\documentclass[12pt]{article}
\usepackage{latexsym,amssymb,amstext,amsmath}
\usepackage{slashed}
\usepackage{mathrsfs}
\usepackage{hyperref}
\usepackage{bbold}
\usepackage{xcolor}
\usepackage{empheq}
\usepackage[toc,page]{appendix}
\numberwithin{equation}{section}
\renewcommand{\Large}{\large} 
\renewcommand{\baselinestretch}{1.2}
\allowdisplaybreaks
\def\slash#1{\rlap{\hbox{$\mskip 1 mu /$}}#1}      
\def\Slash#1{\rlap{\hbox{$\mskip 3 mu /$}}#1}      
\newcommand{\ft}[2]{{\textstyle\frac{#1}{#2}}}

\renewcommand{\theequation}{\thesection.\arabic{equation}}
\renewcommand{\a}{\alpha}
\renewcommand{\b}{\beta}
\newcommand{\g}{\gamma}

\renewcommand{\d}{\delta}

\newcommand{\sH}{{\scriptscriptstyle H}}
\newcommand{\sv}{{\scriptscriptstyle \sf v}}
\newcommand{\ex}{\mathrm{e}}
\newcommand{\im}{\mathrm{i}}

\newcommand {\cC}{{\cal C}}

\newcommand {\cL}{{\cal L}}
\newcommand {\cM}{{\cal M}}

\newcommand {\cR}{{\cal R}}
\newcommand {\cS}{{\cal S}}

\newcommand {\cV}{{\cal V}}

\def\a{\alpha}
\def\b{\beta}

\def\d{\delta}
\def\e{\epsilon}

\def\g{\gamma}

\def\ri{{\rm i}}

\renewcommand{\theequation}{\thesection.\arabic{equation}}

\newcommand{\be}{\begin{equation}}
\newcommand{\ee}{\end{equation}}
\newcommand{\bea}{\begin{eqnarray}}
\newcommand{\eea}{\end{eqnarray}}

\newcommand{\ba}{\begin{array}}
\newcommand{\ea}{\end{array}}

\def\double #1{#1{\hbox{\kern-2pt $#1$}}}

\newcommand{\bsubeq}{\begin{subequations}}
\newcommand{\esubeq}{\end{subequations}}

\newcommand{\veps}{\varepsilon}

\begin{document}

\begin{titlepage}

\begin{center}

\vskip .3in \noindent

{\Large \bf{Higher derivative corrections to BPS black hole attractors \\\vspace{.2cm} in 4d gauged supergravity}}

\bigskip

	Kiril Hristov$^{1}$, Stefanos Katmadas$^{2,3}$ and Ivano Lodato$^{4}$\\

       \bigskip
       $^{1}$Institute for Nuclear Research and Nuclear Energy, Bulgarian Academy of Sciences, \\
        Tsarigradsko Chaussee 72, 1784 Sofia, Bulgaria \\
       \medskip
	   $^{2}$ Dipartimento di Fisica, Universit\`a di Milano--Bicocca, I-20126 Milano, Italy \\
       \medskip
       $^{3}$ INFN, sezione di Milano--Bicocca, I-20126 Milano, Italy \\
       \medskip
       $^{4}$ IISER Pune, Department of Physics, Homi Bhaba Road, Pashan, Pune, India \\

       \vskip .5in
       {\bf Abstract }
       \vskip .1in

       \end{center}

We analyze BPS black hole attractors in 4d gauged supergravity in the presence of higher derivative supersymmetric terms, including a Weyl-squared-type action, and determine the resulting corrections to the Bekenstein-Hawking entropy. The near-horizon geometry AdS$_2 \times$S$^2$ (or other Riemann surface) preserves half of the supercharges in $N=2$ supergravity with Fayet-Iliopoulos gauging. We derive a relation between the entropy and the black hole charges that suggests via AdS/CFT how subleading corrections contribute to the supersymmetric index in the dual microscopic picture.

Depending on the model, the attractors are part of full black hole solutions with different asymptotics, such as Minkowski, AdS$_4$, and hvLif$_4$. We give explicit examples for each of the asymptotic cases and comment on the implications. Among other results, we find that the Weyl-squared terms spoil the exact two-derivative relation to non-BPS asymptotically flat black holes in ungauged supergravity.
\noindent

\vfill
\eject

\end{titlepage}

\tableofcontents

\section{Introduction and summary of results} 
\label{sec:intro}
Although supersymmetric black holes in gauged 4d supergravity are in many respects very similar to their siblings in ungauged supergravity, they have enjoyed considerable attention only in recent years. The first analytic examples were discovered in \cite{Cacciatori:2009iz}, elaborated in \cite{Dall'Agata:2010gj,Hristov:2010ri} and further generalized in various directions in \cite{Klemm:2011xw,Barisch:2011ui,Halmagyi:2013sla,Halmagyi:2013qoa,Katmadas:2014faa,Halmagyi:2014qza} and references therein. These solutions typically have either AdS$_4$ asymptotics or a runaway vacuum that can sometimes be seen as a meaningful solution when embedded in higher dimensions \cite{Hristov:2014eza}, e.g.\ hyperscaling violating Lifshitz \cite{Perlmutter:2012he} (hvLif$_4$) that uplifts to AdS$_5$. The full black hole solutions are quarter-BPS in $N=2$ gauged supergravity, with an enhancement to four supercharges (half-BPS) on the horizon. The horizon can in general have a topology of any Riemann surface, such that the near-horizon geometry is AdS$_2 \times \Sigma_g^2$ with genus $g \geq 0$.

The first example of a microstate counting for these black holes was only achieved recently in \cite{Benini:2015eyy} building on results of \cite{Hristov:2013spa,Benini:2015noa}, where the case of spherical asymptotically AdS$_4$ black holes embeddable in 11d supergravity on S$^7$ \cite{Cvetic:1999xp,Duff:1999gh} was considered. It turned out that the leading macroscopic entropy in this case corresponds to the (partially) twisted index of ABJM theory on S$^1 \times$S$^2$. A similar understanding of the leading black hole entropy also exists for the class of asymptotically hvLif$_4$ black holes discussed in \cite{Hristov:2014eza}, as they can be understood from a dimensional reduction of asymptotically AdS$_5 \times$S$^5$ geometries in string theory and therefore described by a (twisted) compactification of $N=4$ SYM \cite{Benini:2013cda}.

It was further shown in \cite{Hristov:2012nu} that some special gaugings in $N=2$ supergravity lead to a vanishing scalar potential with a non-BPS Minkowski vacuum, where the near-horizon geometry is nevertheless half-BPS and falls in the same class of supersymmetric solutions. It was proven that these solutions are equivalent (at the two derivative level) to the non-BPS black holes in ungauged supergravity \cite{Hristov:2012nu}, therefore suggesting an interesting new string theory point of view \cite{Hristov:2014eba}.

In this work we initiate a study of how higher order derivative terms in the supergravity lagrangian affect the near-horizon geometry AdS$_2 \times \Sigma_g^2$, similar in spirit to the work of \cite{Baggio:2014hua} in 5d gauged supergravity. Building on previous work \cite{deWit:2011gk} in off-shell conformal supergravity\footnote{See also \cite{Barisch-Dick:2013xga} for results based on the entropy function rather than supersymmetry.}, we evaluate the corrections to the macroscopic Bekenstein-Hawking entropy from a four-derivative lagrangian containing the Weyl-squared term. This allows to address several different questions about the above summarized black holes with different asymptotics. We mostly focus on the case of spherical horizon, $\Sigma^2_{g = 0} = S^2$, but also comment on the resulting differences when considering a higher genus Riemann surface $\Sigma^2_{g > 0}$ instead. We use the conformal off-shell formalism \cite{de Wit:1980tn,de Wit:1984px,deWit:1999fp,deWit:2010za,Banerjee,Butter:2013lta,Butter:2014iwa} and give general results about the attractors for all types of higher order derivatives, since our BPS conditions are off-shell. This makes the resulting conditions immediately applicable to the supergravity localization techniques that are being developed in recent years (see e.g.\ \cite{Banerjee:2009af,Dabholkar:2010uh,Murthy:2015zzy} and references therein). The explicit entropy corrections can however be only determined once we choose a particular lagrangian via the formalism of Wald, and we consider a four-derivative supersymmetric lagrangian that includes the $\cC^2$-term, (with $\cC$ the Weyl tensor) obtained in \cite{Bergshoeff:1980is}. We give several explicit examples of models with different asymptotics that we solve completely and obtain some very generic results. We do not relate these results directly with string theory, keeping our higher derivative corrections general within the framework of 4d supergravity and not restricted to ones coming from string compactifications.

Concretely, our main result is deriving a relation between the entropy $\cS$, the central charge $Z$, the central charge of the gauging parameters $L$, and the corrected prepotential $F (X, \hat{A})$ with $F_A \neq 0$ (which together with the gauging parameters defines the lagrangian),
\begin{equation}\label{eq:mainresult}
\cS = - \frac{\pi}{2}\, \mathrm{Im} \left( \frac{Z}{L} + 256\, F_A \right)\ ,
\end{equation}
where the exact definitions and derivation are presented in the main body of the paper. The above equation generalizes the main two-derivative attractor equation, presented first in \cite{Dall'Agata:2010gj}, with higher-derivative corrections entering through the quantities $Z$ and $F_A$ (see also \cite{Amariti:2015ybz}). This relation proved to be of particular importance from a microscopic point of view \cite{Benini:2015eyy} as it suggests how the Witten index of the dual field theory relates to the entropy.

Let us briefly discuss some other interesting results from simple examples with different asymptotics. We leave the considerable amount of technical details behind them for the following sections.
\begin{itemize}

\item Minkowski \\
For the case with Minkowski asymptotics we consider the so-called T$^3$-model with prepotential\footnote{Here $\hat A$ indicates the lowest component of the Weyl-squared chiral multiplet, that generates the higher derivative invariant. See more in the following sections and appendix \ref{B}.}
$$
F = \frac{(X^1)^3}{X^0} + c\,\hat{A}\,\frac{X^1}{X^0}\ ,
$$
and the Fayet-Iliopoulos (FI) gauging parameters $\xi_1 = 0$, $\xi_0 \neq 0$. At two-derivative level ($c=0$) the BPS attractor solution of this model with nonvanishing charges $p^0$ and $q_1$ was shown to be equivalent with the non-BPS attractor of ungauged supergravity. This is due to the vanishing scalar potential, which guarantees that the radii of AdS$_2$ and S$^2$ are equal so that the Ricci scalar vanishes. However, in the gauged model we are considering, we find the following exact ratio between the two radii with higher derivative corrections ($c \neq 0$)
\begin{equation}
\frac{R^2_{S^2}}{R^2_{AdS_2}} = 1 + \frac{192\,c}{p^0\,q_1}\ .
\end{equation}
This shows that the two radii are no longer equal, even though asymptotically we again find a Minkowski vacuum. Therefore we conclude that at the level of Weyl-squared supergravity the apparent equivalence between the half-BPS attractor in gauged supergravity and the non-BPS attractor in ungauged supergravity breaks down. Note however that in a more restricted setting where higher order corrections are directly dictated by string theory one might again find the same equivalence, e.g.\ in this example if there exists a stringy argument that sets $c = 0$ or string theory dictates adding other higher derivative lagrangians, c.f.\ \cite{Butter:2014iwa,Banerjee:2016qvj}.

\item hvLif$_4$ \\
Here we again consider the T$^3$-model as above,
$$F = \frac{(X^1)^3}{X^0} + c\,\hat{A}\,\frac{X^1}{X^0}\ ,$$
but with the orthogonal choice for FI parameters, $\xi_0 = 0$, $\xi_1 \neq 0$. We also choose the orthogonal nonvanishing charges $q_0$ and $p^1$ and note that this model is fully upliftable (for arbitrary value $c$ parametrizing the higher derivative corrections) to 5d gauged supergravity with higher derivative terms. An exact match with 5d results \cite{Baggio:2014hua} cannot be pursued because our lagrangian is only a subclass of the reduced 5d lagrangian, so we leave the comparison for the future. Here we make the following interesting observation. In the two-derivative case ($c=0$) the T$^3$ model with the hvLif asymptotics allows only for higher genus black holes, i.e.\ the near-horizon geometry has an H$_2$ factor or its quotients. This is because T$^3$ model is only a restriction from the general STU model, which instead allows for horizon topologies of all Riemann surfaces. The situation changes completely when we switch on the higher derivative correction, as we find horizon solutions with all possible topologies for suitable values of $c$ even in the T$^3$ model. This signifies the appearance of the so-called {\it small} black holes, i.e.\ black holes of vanishing classical area and corresponding Bekenstein-Hawking entropy that develop a horizon only due to the higher derivative terms. It would be interesting to relate these to the dual description with twisted $N=4$ super Yang-Mills theory \cite{Benini:2013cda,Hristov:2014eza}.

\item AdS$_4$ \\
The AdS$_4$ asymptotics are instead reached in the so-called ``magnetic'' T$^3$ model with prepotential
$$F = \sqrt{X^0 (X^1)^3 + c\,\hat{A}\,X^0\,X^1}\ ,$$
where all FI terms are nonvanishing, $\xi_0 \neq 0$, $\xi_1 \neq 0$, or with the more general STU prepotential in the same magnetic frame. In the $c = 0$ case the $N=2$ two-derivative model can be embedded in maximal 4d gauged supergravity and then to full 11d supergravity on S$^7$. We are unfortunately not aware of any results about the higher derivative lagrangians coming from 11d, therefore our $c \neq 0$ model is not known to have a string theory origin. It is nevertheless interesting to look at the Wald entropy, which takes the general form from \eqref{eq:mainresult} and with a suitable gauge fixing choice (see more in section \ref{subsec:micro} or appendix C of \cite{Benini:2015eyy}) can be further simplified to
\begin{equation}
\cS = - \frac{\pi}{2}\, \mathrm{Im} \left( {\cal R} + 256\, F_A \right)\ .
\end{equation}
In the two-derivative case (when $F_A = 0$ and higher derivative corrections inside the definition of ${\cal R}$ vanish), the quantity ${\cal R}$ corresponds directly to the microscopic result for the Witten index \cite{Benini:2015eyy}. We hope that in future the corrected result above will be also readily comparable with the dual field theory index evaluated with finite $N$ corrections.

\end{itemize}

The rest of this paper is organized as follows. In section \ref{sec:sugra} we give more details about the general conformal supergravity formalism and set up our main conventions. In section \ref{sec:bps} we look particularly at the BPS properties of the near-horizon geometry and simplify the supersymmetry equations to algebraic ones, keeping everything general for any supersymmetric lagrangian. In section \ref{sec:f-term} we specify the F-term lagrangian that we want to work with, discuss the gauge-fixing procedure to on-shell supergravity, and derive the resulting Wald entropy formula. We then formulate the attractor equations in terms of the physical black hole charges as defined from the Maxwell equations, and show how these reproduce exactly the known two-derivative results. Finally, in section \ref{sec:conclusion} we take particular models with different asymptotics and solve explicitly the attractor equations, evaluating the corresponding Wald entropy for each case. We leave some technical details on conformal supergravity for appendix \ref{B}.

\section{Conformal supergravity formalism}
\label{sec:sugra}
Dealing with higher-derivative supergravity is vastly simplified by making use of an off-shell formulation, as the algebra of the various symmetries is fixed, irrespectively of the terms present in the lagrangian. In this paper, we use the formalism of superconformal calculus in order to construct and analyse actions containing terms with four derivatives. In this extended setting, all symmetries are local and act linearly on the various fields, while the Poincare version of the theory can be obtained by gauge fixing the additional symmetries and gauge connections.

The superconformal algebra includes general-coordinate, local Lorentz, dilatation, special conformal, chiral SU(2) and U(1) transformations in the bosonic sector. Its fermionic sector contains the generators of supersymmetry (Q) and special supersymmetry (S) transformations, which square to general-coordinate and special conformal transformations, respectively. The gauge fields associated with general-coordinate transformations (the vielbein $e^a_\mu$), dilatations ($b_\mu$), chiral symmetry ($V_\mu{}^i{}_j$, $ A_\mu$) and Q-supersymmetry ($\psi^i_\mu$) are realized by independent fields. The remaining gauge fields of Lorentz ($\omega^{ab}_\mu$), special conformal ($f^a_\mu$) and S-supersymmetry transformations ($\phi^i_\mu$) are composite fields. The corresponding curvatures and covariant fields are contained in a tensor chiral multiplet, with 24 + 24 off-shell degrees of freedom; in addition to the independent superconformal gauge fields it contains three auxiliary fields: a Majorana spinor doublet $\chi^i$, a scalar $D$ and an anti-selfdual Lorentz tensor $T_{ab}{}^{ij}$ (where $i, j, \dots$ are chiral SU(2) spinor indices). For notational simplicity from now on we indicate this auxiliary ${\rm SU}(2)$ singlets by their ${\rm SU}(2)$ invariant combinations, i.e. $T^-_{ab}=T_{ab}{}^{ij}\,\varepsilon_{ij}$ and $T^+_{ab}=T_{ab\,ij}\,\varepsilon^{ij}$. We refer to the Appendix \ref{B} and \cite{deWit:2010za} for a detailed description of the theory.

In order for the theory to be gauge-equivalent to the Poincare theory, we need to consider additional matter degrees of freedom, the so-called compensating multiplets. The on-shell theory then arises upon solving the equations of motion for the auxiliary fields, $T_{ab}^{\pm}$, $\chi^i$, $D$ and choosing convenient values for the compensating multiplets in order to gauge fix the dilatation, SU(2), U(1) and S-supersymmetry transformations. We will make use of a compensating hypermultiplet, whose scalar fields transform under the SU(2) R-symmetry, and a compensating vector multiplet, which contains a complex scalar field with a nonzero U(1) R-symmetry charge. In the on-shell theory, only the gauge field in the compensating multiplet remains, as the graviphoton, while all other bosonic fields in the compensating multiplets are either frozen by gauge choices or set to zero by the equations of motion of the auxiliary fields.

While this procedure can be carried out explicitly for the two-derivative theory, it becomes rather cumbersome for higher-derivative theories, especially since the equations of motion for auxiliary fields are no longer algebraic. However, one need not follow a gauge fixing procedure but instead simply work in the conformal setting, since any physical results must be invariant under the extra symmetries. This is the point of view we take in most of this paper, while convenient gauge choices are used only to make a comparison with the microscopic results in section \ref{subsec:micro}. An additional advantage of this approach is that the BPS conditions we obtain apply to any off-shell Lagrangian, beyond the particular four-derivative examples discussed here.

We will consider the case of a single hypermultiplet and an arbitrary number of vector multiplets, so that the on-shell theory contains no hypermultiplets and an arbitrary number of vector multiplets. Since we are focused on abelian Fayet-Iliopoulos gauged supergravity, this represents the generic situation. We will not consider any explicit higher-derivative terms for the matter fields, but we will allow for the presence of an arbitrary chiral background superfield \cite{deWit:1996ag}, so that the vector multiplet couplings are controlled by a prepotential, $F(X,\hat A)$, that depends holomorphically on the vector multiplet scalars, $X^I$, and the lowest component scalar field of the chiral multiplet, $\hat A$, and is homogeneous of degree two, as
\begin{equation}\label{eq:prepot-1}
 X^I F_I + 2\, \hat A\, F_A = 2\,F\,.
\end{equation}
Here, $F_I$ and $F_A$ are the derivatives of the prepotential with respect to the $X^I$ and $\hat A$, which have conformal weights 1 and 2 respectively, so that $F(X,\hat A)$ has weight 2. Eventually the chiral multiplet corresponding to $\hat A$ will be identified with a composite chiral multiplet describing the covariant objects of the supergravity multiplet, in order to generate (some) $R^2$-terms in the action, but much of our analysis will not depend on this identification. In fact, the BPS conditions derived in this section are valid for any collection of vector and chiral multiplets, but we find it convenient to use the function \eqref{eq:prepot-1} in order to write the results in a form that can be used directly in the Lagrangian we will choose in the next section.

Our main strategy in this section is to obtain S-invariant BPS conditions, which are valid without choosing any particular gauge for the additional symmetries. We discuss this in some detail in the simple setting of hypermultiplets, defining the relevant linear combinations of fermions giving rise to the S-invariant BPS conditions. In the remaining two subsections we apply the same procedure to the vector multiplets and the Weyl multiplet.

\subsection{Hypermultiplets}
The field content of the hypermultiplets is 4 real scalars, described by the section of an $\mathrm{Sp}(n_\sH)\times\mathrm{Sp}(1)$ bundle, denoted by $A^\alpha_i$, and 2 fermions $\zeta^{\alpha}$, where the indices $\alpha,\beta \dots \!= 1, \dots,\, 2\,n_\sH$ for $n_\sH$ hypermultiplets. One can define a covariantly constant hermitian tensor $G^{\alpha\bar\beta}$ (which is used in raising and lowering indices) and of a covariantly constant skew-symmetric tensor $\Omega^{\alpha\beta}$ (and its complex conjugate
$\bar\Omega^{\bar\alpha\bar\beta}$ satisfying
$\Omega_{\bar\alpha\bar\gamma}\bar\Omega^{\bar\beta\bar\gamma}= \delta_{\bar\alpha}{}^{\bar\beta}$).
These can be used to define the pseudo-reality condition on the section $A_i{}^\alpha$, through the constraint
\begin{equation}
\varepsilon^{ij} \bar\Omega^{\bar\alpha\bar\beta} G_{\bar\beta\gamma}A_j{}^\gamma = A^i{}^{\bar\alpha} \equiv
(A_i{}^\alpha)^\ast\,,
\end{equation}
where $\varepsilon^{ij}$ is the antisymmetric symbol for the SU(2) indices. Furthermore, one can define the hyper-K\"ahler potential,
\begin{align}\label{eq:pot-H}
  \chi_\sH =&\, \tfrac12\,\varepsilon^{ij}\bar\Omega_{\alpha\beta} A_i{}^{\alpha}\,A_j{}^{\beta}  \,,
\end{align}
which characterises the geometry of the target space.

The supersymmetry variations for hypermultiplets read
\begin{eqnarray}
  \label{eq:4D-hypertransf}
\d A_i{}^\a &=&
2\, \bar\e_i\zeta^\a
+ 2\, \varepsilon_{ij} \,G^{\a\bar\b}\Omega_{\bar\b\bar\gamma} \, \bar\e^j\zeta^{\bar\gamma} \,,
\nonumber \\
\d \zeta^\a &=& \Slash{D} \,A_i{}^\a \e^i + 2\,\xi_I X^I \, t^{\a}{}_{\b} A_i{}^{\b}\varepsilon^{ij}\epsilon_j
 + A_i{}^\a\,\eta^i\,,
\end{eqnarray}
where the spinors $\e^i$, $\eta^i$ stand for the Q- and S- supersymmetry parameters respectively. Note that we have included a coupling to vector multiplets, through a gauging described by the constants, $\xi_I$, and an anti-hermitian generator $t^{\a}{}_{\b}$. This is reflected in the covariant derivative, $D_\mu$, which contains covariantization with respect to all superconformal symmetries and the gauge symmetries of $A_i{}^{\a}$, defined as
\begin{align}
 D_\mu\,A_i{}^\a = &\, {\cal D}_\mu \,A_i{}^\a + \text{fermions}
 \nonumber\\
 =&\, \partial_\mu\,A_i{}^\a + b_\mu A_i{}^\a + \tfrac12\,{\cal V}_\mu{}_i{}^j\,A_j{}^\a  - \xi_I W_\mu^I \, t^{\a}{}_{\b} A_i{}^{\b} + \text{fermions} \,.
\end{align}
Here, the $W_\mu^I$ are gauge fields belonging to vector multiplets with scalar components given by the $X^I$, for $I,\,J\dots = 1,\dots,n_\sv$. The constants $\xi_I$ are the so-called Fayet-Iliopoulos (FI) gauging parameters appearing in standard literature. We suppress the gauge coupling constant $g$, noting that one needs to send all parameters $\xi_I$ to zero to get back to the ungauged case.

Inspired by the structure of \eqref{eq:pot-H} one can define a spinor that transforms into a constant under S-supersymmetry, as
\begin{align}\label{eq:Zeta-H}
  \zeta^\mathrm{H}_i  =&\,\chi_\mathrm{H}^{-1}  \bar\Omega_{\alpha\beta} A_i{}^\alpha\, \zeta^\beta \,,
\\
\label{eq:Zeta-H-var}
  \delta\zeta^\mathrm{H}_i = &\, \tfrac12
\slash{k}\,\,\varepsilon_{ij}\, \epsilon^j + \slash{k}\,_{ij}\,\epsilon^j
+2\,X^I  \mu_{I}{}_{ij} \varepsilon^{j k} \, \epsilon_k + \varepsilon_{ij} \eta^j \,,
\end{align}
so that it may be used as a compensating spinor in the construction of S-invariant combinations. Here, we suppress the terms proportional to fermion fields, as they are not relevant for deriving BPS conditions for bosonic fields. The moment maps, $\mu_I{}_{ij}$, are defined as
\begin{equation}\label{eq:mu-def}
 \mu_I{}_{ij} = \chi_\sH^{-1} \xi_I \,\left( A_i{}^{\a} \bar\Omega_{\alpha\beta} t^\b{}_{\gamma} A_j{}^{\gamma} \right)\,,
 \qquad
 \mu_I{}^{ij} \equiv ( \mu_I{}_{ij} )^* = \varepsilon^{ik}\varepsilon^{jl}\mu_I{}_{kl}\,,
\end{equation}
where the pseudo-reality condition can be shown using the properties listed in \cite{deWit:1999fp}, for instance. Similarly, the vectors $k_\mu$ and $k_{\mu}{}^i{}_{j}$ are defined as the singlet and triplet in the decomposition of the scale-invariant combination
\begin{equation}
-\chi_\sH^{-1} \bar\Omega_{\alpha\beta} \varepsilon^{ik} A_k{}^{\alpha} D_\mu A_j{}^{\beta} =
\tfrac12\, k_\mu \,\d^i_{j} + k_{\mu}{}^i{}_{j}\,,
\end{equation}
where
\begin{align}\label{eq:k-hyp-ex}
k_\mu =&\, \chi_\sH^{-1} {\cal D}_\mu \chi_\sH \,,
\end{align}
while the explicit form of the triplet, $k^i{}_j=-\varepsilon^{ik}\, k_{kj}$, will not be relevant in what follows.

We can now consider the following S-invariant variation of the hyperinos,
\begin{align}
\label{eq:hyper_var}
 \delta(\zeta^\alpha+\varepsilon^{ij}A_i{}^\alpha\,\zeta^\mathrm{H}{}_j)
 =&\, \chi_{\sH}^{1/2}\Slash{D}\left(\chi_{H}^{-1/2}  A_i{}^{\alpha}\right)\,\epsilon^i
- A_i{}^{\alpha}\,\slash{k}{}^i{}_{j}\,\epsilon^j
\nonumber \\
&\,
+2\, X^I \,
\left( \xi_I\,t^{\a}{}_{\b} A_i{}^{\b}
+ \varepsilon^{kl} A_k{}^{\alpha} \mu_I{}_{li} \right)\varepsilon^{ij}\epsilon_j
\,.
\end{align}
This final form of the variation can be used to derive BPS conditions, given a particular ansatz for the geometry and the amount of supersymmetry. Assuming full supersymmetry, all terms are linearly independent and must therefore vanish separately, leading to the classification of \cite{deWit:2011gk}. Alternatively, one may assume a particular spacetime ansatz and derive the possible BPS classes of solutions allowed by \eqref{eq:hyper_var} and the analogous conditions arising from other multiplets \cite{deWit:2011gk}.

In the next section we will consider a 1/2-BPS ansatz for AdS$_2\times$S$^2$, in the presence of a single (compensating) hypermultiplet, so that one can set $n=1$ above and the indices $\a$, $\b=1,2$. It follows that we can take $\bar\Omega_{\alpha \beta}$ to be the corresponding antisymmetric symbol for these indices and that the generator $t^{\a}{}_{\b} \in$U(1) as a subgroup of SU(2). We assume these restrictions for the remainder of the paper, noting that the decomposition \eqref{eq:hyper_var} does not rely on this simplification.

\subsection{Vector multiplets}
The field content of a vector multiplet is a complex scalar, $X$, an SU(2) doublet of fermions, $\Omega_i$, a gauge field, $W_{\mu}$, described by its field strength, $F_{\mu\nu}$, and an SU(2) triplet of auxiliary scalars, $Y_{ij}$. We will consider an arbitrary number, $n_\sv$, of vector multiplets, labelled by an index $I,\,J\dots = 1,\dots,n_\sv$.
The Q- and S-supersymmetry transformations for the vector multiplet
take the form,
\begin{align}
  \label{eq:variations-vect-mult}
  \delta X =&\, \bar{\epsilon}^i\Omega_i \,,\nonumber\\
  \delta\Omega_i =&\, 2 \Slash{D} X\epsilon_i
     +\ft12 \varepsilon_{ij} \hat{F}_{\mu\nu}
   \gamma^{\mu\nu}\epsilon^j +Y_{ij} \epsilon^j
     +2X\eta_i\,,\nonumber\\
  \delta A_{\mu} = &\, \varepsilon^{ij} \bar{\epsilon}_i
  (\gamma_{\mu} \Omega_j+2\,\psi_{\mu j} X)
  + \varepsilon_{ij}
  \bar{\epsilon}^i (\gamma_{\mu} \Omega^{j} +2\,\psi_\mu{}^j
  \bar X)\,,\nonumber\\
\delta Y_{ij}  = &\, 2\, \bar{\epsilon}_{(i}
  \Slash{D}\Omega_{j)} + 2\, \varepsilon_{ik}
  \varepsilon_{jl}\, \bar{\epsilon}^{(k} \Slash{D}\Omega^{l)
  } \,,
\end{align}
where $\psi_\mu{}^j$ stands for the gravitini and we use the convenient shorthand
\begin{equation}
\hat{F}_{\mu\nu} = F^{-}_{\mu\nu} -\ft 14\, \bar X\, T^-_{\mu\nu}\,,
\end{equation}
which appears frequently below.

A vector multiplet is a special case of a scalar chiral multiplet, which is the most general multiplet depending on half of the spinorial variables in superspace. Chiral multiplets are characterised by the conformal weight, $w$, of their lowest component, a complex scalar, $A$, generalising the vector multiplet scalar, for which $w=1$. The higher components include two doublets of spinors, $\Psi_i$, $\Lambda_i$, an anti-selfdual tensor, $G_{ab}^-$, a complex SU(2) triplet of scalars, $B_{ij}$, and an additional complex scalar, $C$, with the supersymmetry transformations of the first components given by
\begin{align}
  \label{eq:conformal-chiral}
  \delta A =&\,\bar\epsilon^i\Psi_i\,, \nonumber\\[.2ex]
  \delta \Psi_i =&\,2\,\Slash{D} A\epsilon_i + B_{ij}\,\epsilon^j +
  \tfrac12   \gamma^{ab} G_{ab}^- \,\varepsilon_{ij} \epsilon^j + 2\,w
  A\,\eta_i\,.
\end{align}
Since we will be using only a particular composite chiral multiplet, made out of the Weyl multiplet, we need not consider the conditions arising from independent chiral multiplets, as they will be automatically satisfied once the BPS conditions for the Weyl multiplet are imposed. However, the variations in \eqref{eq:conformal-chiral} will be useful below, since the coupling of the vector multiplets is controlled by the prepotential $F(X^I,\hat A)$ in \eqref{eq:prepot-1}. While the use of a particular function related to the couplings in the Lagrangian we intent to use is not necessary at this point, it is very convenient for later developments, so we consider the dependence on the chiral background through $\hat{A}$ already in this section.

We thus proceed to define the function
\begin{align}
  \label{eq:pots}
  \ex^{-K} =&\, \mathrm{i}\,(\bar X^I F_{I}-X^I \bar F_{I}) \,,
\end{align}
and the fermion
\begin{align}
  \label{eq:Omega-V}
  \Omega^\mathrm{\sf v}_i=&\,
  \frac{\rm i}{2}\,\ex^{K} \Big[\left(\bar X^I F_{IJ}  - \bar F_J\right) \Omega_i{}^J +\,\bar{X}^I F_{A\,I} \hat \Psi_i \Big]   \,,
\nonumber\\
\delta\Omega^\mathrm{\sf v}_i  =&
\tfrac12 \,\ex^{K} \Slash{D} \,\ex^{-K} \epsilon_i
 + i\Slash{\mathcal{A}} \,\, \epsilon_i
 + \tfrac14\,\varepsilon_{ij} \, {\cal F}_{ab}^- \, \gamma^{ab} \epsilon^j
 +\tfrac12\,{\cal Y}_{ij} \, \epsilon^j
 + \eta_i\,,
\end{align}
where we ignored higher-order fermionic terms in the variation.
The quantity ${\cal A}_\mu$ resembles a covariantized (real) K\"ahler
connection, while ${\cal F}_{ab}^-$ and ${\cal Y}_{ij}$ are an anti-selfdual tensor and a complex $SU(2)$ triplet respectively,
\begin{eqnarray}
{\cal A}_\mu &=& \tfrac12\,\ex^{K}\,
\Big( \bar X^I \stackrel{\leftrightarrow}{\cal D}_\mu  F_I
-\bar F_I \stackrel{\leftrightarrow}{\cal D}_\mu X^{I} \Big)
\nonumber \\
{\cal F}_{ab}^- &=&
\mathrm{i}\,\ex^{K}(\bar X^I F_{IJ} - \bar F_{J}) \,\hat F^J_{ab}
+\mathrm{i}\,\ex^{K}\bar X^I F_{{A}I}\hat G_{ab}  \;\;.
\label{curlyf}
\nonumber\\
{\cal Y}_{ij} &=&
\mathrm{i}\,\ex^{K}(\bar X^I F_{IJ} - \bar F_{J}) \,Y^J_{ij}
+\mathrm{i}\,\ex^{K}\bar X^I F_{{A}I}\hat B_{ij}  \;\;.
\label{curlyY}
\end{eqnarray}

We now consider the supersymmetry variations of the S-invariant spinors one can construct from the vector multiplet fermions and the hypermultiplet compensator $\zeta^{\sH}{}_i$ in \eqref{eq:Zeta-H}, which read
\begin{align}
\label{eq:vect_vect_var}
\delta\left( \Omega_i{}^I -2\,X^I\, \Omega^{\sf v}_i\right)= &\,
2\,\ex^{-K/2} \Slash{D}\left(\ex^{K/2} X^I\right)\epsilon_i
-2 \im\,X^I\Slash{\mathcal{A}} \,\, \epsilon_i
\nonumber\\
&\,
+\tfrac12\, \varepsilon_{ij} \left( \hat F^I_{\mu\nu}{}^- - X^I {\cal F}_{\mu\nu}^- \right) \gamma^{\mu\nu}\epsilon^j
+ \left( Y^I_{ij} - X^I\,{\cal Y}_{ij} \right) \epsilon^j\ ,
\\
\label{eq:vect_hyper_var}
\delta\left(\Omega^\mathrm{V}_i +\varepsilon_{ij}\zeta^\mathrm{H}{}^j\right) = &i\Slash{\mathcal{A}} \,\, \epsilon_i
-\tfrac12 \Slash{\mathcal{D}} \log \left( e^K \chi_\sH\,\right) \epsilon_i + \varepsilon_{ij}\slash{k}\,^{jk}\,\epsilon_k
 \nonumber \\
&
 + \tfrac14\,\varepsilon_{ij} \, {\cal F}_{ab}^- \, \gamma^{ab} \epsilon^j
 +\tfrac12\,{\cal Y}_{ij} \, \epsilon^j
 + 2 \,\bar{X}^I \varepsilon_{i j}  \mu_I{}^{j k} \varepsilon_{k l} \, \epsilon^l \,,
\end{align}
where we used \eqref{eq:k-hyp-ex}. This is the final form for the supersymmetry variations, without imposing any restrictions on the spinor parameters. Note that the variables $\ex^{K/2} X^I$ and $\ex^{K} \chi_\sH$ are scale invariant, so that they represent natural variables to be used in physical quantities.

\subsection{The Weyl multiplet}
The covariant fields of the Weyl multiplet comprise the field strengths for the various gauge fields and the auxiliary fields of the multiplet. In the fermionic sector, this amounts to the field strength of the gravitino, $R(Q)_{ab}^i$, the corresponding field strength for the S-supersymmetry gauge field $\phi_\mu$ and the auxiliary spinor $\chi^i$. In view of the conventional constraint, $\gamma^\mu R(Q)_{\mu \nu}{}^i = - \tfrac32\, \gamma_{\nu}\chi^i$, both $\phi_\mu$ and $\chi^i$ are given in terms of the gravitino, so that one only need consider BPS conditions arising from variations of its field strength, $R(Q)_{ab}^i$. In addition, the derivative of at least one covariant fermion must be considered, in order to ensure that the variations of all derivatives of fermionic fields vanish \cite{LopesCardoso:2000qm}. This imposes a constraint on the fields of the Weyl multiplet, irrespectively of the type of fermion chosen, so we take it to be the derivative of the compensating fermion, $\zeta^\sH{}_i$, for simplicity.

We now give the Q-variations of the relevant S-invariant combinations one can built out of $R(Q)_{ab}{}^i$ and $D_\mu\zeta^{\mathrm{H}}{}_i$, using the compensating fermion, $\zeta^{\mathrm{H}}{}_i$, defined in \eqref{eq:Zeta-H}. The variation of the gravitino field strength reads
\begin{align}
  & \delta \left( R(Q)_{ab}{}^i
   -\tfrac1{16}\,T_{cd}^{-}\,
   \gamma^{cd} \gamma_{ab} \, \zeta^\mathrm{H}{}^i\right)
\nonumber\\[.2ex]
 &=\,- \tfrac14 \chi_H^{1/2} \Slash{D}(\chi_H^{-1/2}T_{ab}^-) \,\varepsilon^{ij}\epsilon_j
   +\tfrac14\,\left( T_{d[a}^-\,k_{b]} - k^c T_{c[a}^-\,\eta_{b]d} \right) \gamma^{d}\,\varepsilon^{ij}\epsilon_j
\nonumber\\[.2ex]
&\,\quad +   R({\cal V})^-{}_{\!\!ab}{}^i{}_j \, \epsilon^j
    - \tfrac12 \hat{R}(M)_{ab}{}^{\!cd}\, \gamma_{cd} \epsilon^i
   -\tfrac1{8}\,T^-_{cd}\,\varepsilon^{ij}\, \bar{X}^I  \mu_I{}_{j k} \, \gamma^{cd} \gamma_{ab} \,\epsilon^k \,,
   \label{eq:RQ-var}
\end{align}
where we used \eqref{eq:k-hyp-ex} to form the appropriate scale-invariant combination $\chi_\sH^{-1/2}T_{ab}{}^-$ under the covariant derivative.
The supersymmetry variation of the derivative $D_\mu \zeta^{\mathrm{H}}{}_i$ reads
\begin{align}\label{eq:dzeta-var-0}
\delta D_\mu \zeta^\mathrm{H}_i = &\
f_\mu^a\gamma_a\varepsilon_{ij} \epsilon^j
- \tfrac18\,\varepsilon_{ij}\, R({\cal V})^+{}_{\!\!ab}{}^j{}_k\,\gamma^{ab}\gamma_{\mu}\epsilon^k
- \tfrac{\mathrm i}4\, R(A)^+{}_{\!\!ab}\,\gamma^{ab}\gamma_{\mu}\epsilon_i
\nonumber\\
&\,
-\tfrac{1}{16}\,T_{ab}^{+}\,\gamma^{ab}\gamma_{\mu}\mu_I{}_{i j} X^I \epsilon^j + \mu_I{}_{ij} X^I \gamma_{\mu}\varepsilon^{jk} \eta_k
\,,
\end{align}
up to terms proportional to derivatives of scalar fields and the tensor $T_{ab}{}^-$. As will be shown in the next section, all such derivatives vanish for 1/2-BPS backgrounds, once all the previous BPS conditions are imposed, so that we restrict to this case for brevity. The supersymmetry variation of the S-invariant combination then reads
\begin{align}\label{eq:dzeta-var}
\delta \left( D_\mu \zeta^\mathrm{H}_i - \mu_I{}_{ij} X^I \gamma_{\mu}\zeta^\mathrm{H}{}^j \right) =&\
f_\mu^a\gamma_a\varepsilon_{ij} \epsilon^j
- \tfrac18\,\varepsilon_{ij}\, R({\cal V})^+{}_{\!\!ab}{}^j{}_k\,\gamma^{ab}\gamma_{\mu}\epsilon^k
- \tfrac{\mathrm i}4\, R(A)^+{}_{\!\!ab}\,\gamma^{ab}\gamma_{\mu}\epsilon_i
\nonumber\\
&\,
-\tfrac{1}{16}\,T_{ab}^{+}\,\gamma^{ab}\gamma_{\mu}\mu_I{}_{i j} X^I \epsilon^j
-2\, X^I \mu_I{}_{ij}\,\bar{X}^J \mu_J{}^{jk} \gamma_{\mu}\varepsilon_{kl} \epsilon^l \,.
\end{align}
Note the presence of a bare K-boost gauge field, $f_\mu^a$, in \eqref{eq:dzeta-var-0}-\eqref{eq:dzeta-var}, originating in the inhomogeneous S-supersymmetry transformation of the compensating fermion in \eqref{eq:Zeta-H-var}. It follows that \eqref{eq:dzeta-var} is a constraint on the fields of the Weyl multiplet, ensuring the vanishing of all fermionic derivatives.

\section{Attractor ansatz and supersymmetry}
\label{sec:bps}

In this section we consider the conditions arising from supersymmetry, assumming a 1/2-BPS condition. We focus on systems containing a single, compensating, hypermultiplet and an arbitrary number of vector multiplets, allowing for an abelian gauging of the hypermultiplet using the vector fields in the vector multiplets.

In order to proceed, we adopt a particular projection for the Killing spinor preserved by the background, which breaks the SU(2) R-symmetry invariance down to a U(1). In particular, we choose the supersymmetry parameter to be such that \begin{align}\label{eq:KS-ansatz}
\gamma^{23}\epsilon^i= i\, \sigma_{3\;\;j}^{\;\;i} \,\epsilon^j\,,
\end{align}
where $\sigma_3$ stands for the third Pauli matrix. One can straightforwardly replace this by any element of the SU(2) algebra, but we restrict to this choice for simplicity, without loss of generality.

Note that a 1/2-BPS Killing spinor need not satisfy a projection of the type \eqref{eq:KS-ansatz} in general. However, in the case of a single gauged hypermultiplet, the constant SU(2) element parametrising the gauging must be compatible with the properties of the spinor, as will be seen below. Conversely, it has been shown that \eqref{eq:KS-ansatz} is the only possible condition on a spinor, assuming an AdS$_2\times$S$^2$ background \cite{deWit:2011gk}. These are the backgrounds we are interested in, so we will adopt this choice henceforth. This choice for the Killing spinor projection is eventually equivalent to choosing the matrix $t^{\alpha}{}_{\beta}$ controlling the gauging in \eqref{eq:mu-def} to be
\begin{align}\label{eq:mu_I}
t^{\beta}{}_{\gamma} \sim \sigma_3{}^{\beta}{}_{\gamma}\ ,
\quad \Rightarrow \quad
\mu_I{}_{ij} = {\rm i}\, \xi_I\, \varepsilon_{i k}\, \sigma_3{}^k{}_j\ ,
\end{align}
as will be explained in details shortly. This identity is not immediately needed here in the off-shell context but the reader familiar with Poincare supergravity will recognize more easily the physical FI parameters $\xi_I$ appearing this way.

\subsection{BPS conditions on the scalars}\label{sec:Scal-BPS}
It is important to note that \eqref{eq:KS-ansatz} only allows to reduce terms involving exactly two gamma matrices, so that terms appearing in the various spinor variations can be meaningfully separated in those containing even and odd numbers of gamma matrices. It then follows that each group of terms must vanish separately.

We therefore first consider the terms with an odd number of gamma matrices in \eqref{eq:hyper_var} and \eqref{eq:vect_vect_var}-\eqref{eq:vect_hyper_var}, setting to zero the terms transforming in different SU(2) representations. We thus find the conditions
\begin{gather}
{\cal D}_\mu\left(\chi_{H}^{-1/2}  A_i{}^{\alpha}\right)=0\,,
\label{eq:hyper-con}
\\
{\cal D}_\mu\left(\ex^{K/2} X^I\right)=0\,,
\label{eq:vec-con}
\\
{\cal D}_\mu\left(\ex^{-K} \chi^{-1}_\sH \right) = \mathcal{A}_\mu = k_\mu{}^i{}_j= 0\,,
\label{eq:vec-hyp-con}
\end{gather}
which are both conformally and K-invariant. Further restrictions arise by considering intergrability conditions on these equations, explicitly
\begin{align}
 {\cal D}_{[\mu}{\cal D}_{\nu]}\left(\chi_{H}^{-1/2}  A_i{}^{\alpha}\right)=0
 &\,\quad\Rightarrow\quad
 \tfrac12\, R({\cal V})_{\mu\nu}{}_i{}^jA_j{}^{\alpha} - \xi_I \, t^{\a}{}_{\b} A_i{}^{\b}\,F_{\mu\nu}^I=0\,,
 \label{eq:RV-con}
\\
 {\cal D}_{[\mu}{\cal D}_{\nu]}\left(\ex^{K/2} X^I\right)=0
 &\,\quad\Rightarrow\quad
 R(A)_{\mu\nu} =0\,.
\end{align}
Since the field strength of the U(1) R-symmetry gauge field vanishes, $A_\mu$ is locally vanishing everywhere, so it drops out of all covariant derivatives. Consequently, we solve \eqref{eq:hyper-con} and \eqref{eq:RV-con} by imposing that the corresponding gauge fields satisfy the condition
\begin{equation}\label{eq:V}
 \tfrac12\, {\cal V}_{\mu}{}_i{}^j = \mu_{I\,ik}\varepsilon^{kj} W_\mu^I \,,
\end{equation}
and that the rescaled section $\chi_{H}^{-1/2}  A_i{}^{\alpha}$ is constant, as is $\mu_I{}_{ij}$.

The above results imply that one can do convenient gauge choices for some of the superconformal symmetries, in order to simplify the following discussion. Note that all conditions \eqref{eq:hyper-con}-\eqref{eq:vec-hyp-con} involve derivatives of scale invariant combinations, so that the spacetime dependence of all scalar fields can be restricted to a single function, which can be taken to be $\ex^{-K}$. Using a conformal transformation, we can set this function to a constant, thus reducing all scalar fields in the hyper- and vector multiplets to constants. Note that this still leaves a residual rigid conformal symmetry, which is unphysical and must therefore drop out from all physical quantities.
With this gauge choice, we obtain that $k_\mu$, defined in \eqref{eq:k-hyp-ex} vanishes, so that setting to zero the terms with an odd number of gamma matrices in \eqref{eq:RQ-var}, leads to the condition
\begin{equation}
{\cal D}_\mu(\chi_\sH^{-1/2}T^-_{ab})=0\,.
\end{equation}
This establishes that a covariantly constant anti-selfdual tensor can be defined on the background.

Additionally, we can make a similar gauge choice for the hypermultiplet sections exploiting the $\mathrm{SU}(2)$ gauge symmetry. For a general situation involving many hypermultiplets this is not possible, but in this paper we are interested in the case of a single hypermultiplet, whose degrees of freedom can be gauged away to obtain an on-shell theory. For a single hypermultiplet, we can use an $\mathrm{SU}(2)$ rotation to fix the constant $\chi_{H}^{-1/2}  A_i{}^{\alpha}$ as
\begin{align}\label{eq:gauge-fix-chi}
 \chi_{H}^{-1/2}  A_i{}^{\alpha} = \delta_i^{\alpha}\ ,
\end{align}
which can be used to identify the indices $\alpha, \beta\dots =1,2$ with the $\mathrm{SU}(2)$ indices $i,j\dots$. We can then write $\bar\Omega_{ij} = \varepsilon_{ij}$ and rewrite \eqref{eq:mu-def} as
\begin{align}
\label{eq:gauge-fix-generator}
 \mu_{i j, I} = \xi_I \varepsilon_{i k} t^k{}_{j}\ ,
\end{align}
which will be used in the rest of this paper to translate the gauging $\mu_{i j, I}$ to the FI terms. As will be seen in subsection \ref{sec:Sol-BPS}, the generator $t^i{}_{j}$ will be identified with $\mathrm{i}\, \sigma^3{}^i{}_j$ by consistency.

\subsection{Bosonic background}
We are looking for static spherical black hole attractor geometries with constant scalars as shown above, and introduce the following notation for the AdS$_2 \times$S$^2$ metric\footnote{It is straightforward to generalize our results to toroidal and higher genus horizons, but for most of the discussion here and later in the paper we look at the spherical case. We come back to general horizon topologies in \ref{subsec:finalbps} and in some of the explicit examples we show at the end.}:
\begin{equation}
  \label{eq:ads2xs2}
    \mathrm{d}s^2 = g_{\mu\nu} \mathrm{d}x^\mu\mathrm{d}x^\nu =
  v_1\Big(-r^2\,\mathrm{d}t^2 + \frac{\mathrm{d} r^2}{r^2}\Big)
  + v_2 \Big(\mathrm{d} \theta^2 +\sin^2\theta
  \,\mathrm{d}\varphi^2\Big)\,,
\end{equation}
whose non-vanishing Riemann curvature components are equal to
\begin{equation}
  \label{eq:geo}
  R_{\underline{a} \underline{b}}{}^{\underline{c} \underline{d}}  =2\, v_1^{-1}
  \delta_{\underline{a}\underline{b}}{}^{\underline{c}\underline{d}}
  \,,\qquad
  R_{\hat a \hat b}{}^{\hat c \hat d} =-2 \,v_2^{-1} \delta_{\hat
    a\hat b}{}^{\hat c\hat d} \,,
\end{equation}
so that the four-dimensional Ricci scalar equals $R= 2
(v_1^{-1}-v_2^{-1})$. Observe that we used tangent-space indices
above, where $\underline{a}, \underline{b}, \ldots$ label the flat
$\mathrm{AdS}_2$ indices $(0,1)$ associated with $(t,r)$, and $\hat a,
\hat b, \ldots$ label the flat $S^2$ indices $(2,3)$ associated with
$(\theta,\varphi)$. The non-vanishing components of the
auxiliary tensor field are parametrized by a complex scalar $w$,
\begin{equation}
  \label{eq:T-into-w}
  T^-_{01}=\mathrm{i}\,T^-_{23}= -\,w\,.
\end{equation}
Using these paramterizations one finds the following expressions for the
bosonic part of the special conformal gauge field $f_a{}^b$ (see the appendix for more details),
\begin{align}
  \label{eq:K-gaugefield}
  f_{\underline{a}}{}^{\underline{b}}=&\, \big(\frac16 (2\,v_1^{-1}+
  v_2^{-1})  - \frac14
  D -\frac1{32} |w|^2\big) \delta_{\underline{a}}{}^{\underline{b}} +\frac12
  R(A)_{23}\, \varepsilon_{\underline{a}}{}^{\underline{b}}\,,
  \nonumber\\
  f_{\hat a}{}^{\hat b}=&\, \big(-\frac16 (v_1^{-1} +2\,v_2^{-1}) -
  \frac14D +\frac1{32} |w|^2\big )\delta_{\hat a}{}^{\hat b}+
  \frac12 R(A)_{01}\,\varepsilon_{\hat a}{}^{\hat b}\,,
\end{align}
where the two-dimensional Levi-Civita symbols are normalized by
$\varepsilon^{01}=\varepsilon^{23} =1$. The non-zero components of
the modified curvature $\mathcal{R}(M)_{ab}{}^{cd}$ are given by,
\begin{align}
  \label{eq:R(M)-values}
  \mathcal{R}(M)_{\underline{a} \underline{b}}{}^{\underline{c}
    \underline{d}} =&\,(D+\frac13 R)\, \delta_{\underline{a}\underline{b}}
  {}^{\underline{c}\underline{d}}  \,,\nonumber\\
  \mathcal{R}(M)_{\hat a \hat b}{}^{\hat c \hat d} =&\,(D+\frac13 R)\,
  \delta_{\hat a\hat  b}{}^{\hat c\hat d} \,,\nonumber\\
  \mathcal{R}(M)_{\underline{a} \hat b}{}^{\underline{c} \hat d}
  =&\,\frac12(D-\frac16 R)\, \delta_{\underline{a}}{}^{\underline{c}} \,
  \delta_{\hat b}{}^{\hat d} - \frac12
  R(A)_{23}\,\varepsilon_{\underline{a}}{}^{\underline{c}}\,
  \delta_{\hat b}{}^{\hat d}-\frac12
  R(A)_{01}\,\delta_{\underline{a}}{}^{\underline{c}}\, \varepsilon_{\hat
    b}{}^{\hat d}\, .
\end{align}
We refer to the appendix for the general definitions of these quantities, which appear in the
superconformal transformation rules of the Weyl multiplet fields and
are therefore needed below.

\subsection{Solving the BPS equations}\label{sec:Sol-BPS}
Proceeding with solving the BPS conditions, we start with \eqref{eq:vect_vect_var} and \eqref{eq:vect_hyper_var}, where the first line of each has already been analysed in subsection \ref{sec:Scal-BPS}. The second lines in each of \eqref{eq:vect_vect_var} and \eqref{eq:vect_hyper_var} are even in the gamma matrices, so we need to use the projection \eqref{eq:KS-ansatz} in order to obtain conditions on the fields. Given spherical symmetry, we can assume that only the $(0,1)$ and $(2,3)$ components of the field strengths are non-zero. Under this assumption, the condition from \eqref{eq:vect_vect_var} takes the form
\begin{align}
\label{eq:vectorBPS}
2\,{\rm i}\varepsilon_{ij}\Big(\hat F_{23}^{I\,-}- X^I \mathcal{F}^-_{23}\Big)\sigma_{3\;\;k}^{\;\;j} + \left( Y^I_{ij} - X^I\,{\cal Y}_{ij} \right)\delta^j_k = 0\,,
\end{align}
where we used the anti-selfduality of both $\hat F^{I\,-}_{ab}$ and $\mathcal{F}_{ab}$.
From \eqref{eq:vect_hyper_var} we also obtain the condition:
\begin{align}
\label{eq:scalarBPS}
{\rm i}\varepsilon_{ij}\,\mathcal{F}^-_{23}\sigma_{3\;\;k}^{\;\;j}  +\,\tfrac12{\cal Y}_{ij} \delta^j_k
-2\,\bar{X}^I\, \mu_{I\,ij} \delta^j_k = 0\ ,
\end{align}
but notice that \eqref{eq:scalarBPS} is implied by \eqref{eq:vectorBPS} upon contraction with the sections, therefore it is not an independent constraint.

Imposing spherical symmetry for the curvature $R({\cal V})^-{}_{ab}{}^i{}_j$, we find
\begin{align}
2\,{\rm i} R({\cal V})^-{}_{23}{}^i{}_j + 3\, D\,\sigma_3{}^i{}_j -{\rm i} T^-_{23}\bar{X}^I \mu_I{}^{ik}\,\varepsilon_{kj}= 0\,.
\end{align}
Returning to \eqref{eq:RQ-var}, the terms with an even number of gamma matrices lead to the conditions
\begin{align}
  \label{eq:RV-A-epsilon}
  \mathrm{i}R(\mathcal{V})^-_{23}{}^i{}_j =&\,-(D+\tfrac1{12} R)\,\sigma_3{}^i{}_j \,,\nonumber\\
  w\, \bar{X}^I \mu_I{}^{ik}\varepsilon_{kj} =&\, -(D-\tfrac16 R)\,\sigma_3{}^i{}_j \,.
\end{align}
The terms with an even number of gamma matrices in \eqref{eq:dzeta-var} then yield
\begin{align}
  \label{eq:RV-A-epsilon-cc}
  \mathrm{i}R(\mathcal{V})^+_{23}{}^i{}_k \sigma_3{}^k{}_j =&\,
  \tfrac12(v_1^{-1}+ v_2^{-1}-\tfrac14|w|^2)
  \,,\nonumber\\
  \bar w\, \varepsilon^{ik} X^I\mu_{I\,kj} \, =&\, -\tfrac18 |w|^2 \,\sigma_3{}^i{}_j \,.
\end{align}
Combining these equations leads to,
\begin{align}
  \label{eq:generic}
 \bar w\, \varepsilon^{ik} X^I\mu_{I\,kj} =&\,  -w\, \bar{X}^I\mu_{I}{}^{ik}\varepsilon_{kj}
 \quad\Rightarrow\quad
 \left( \bar w\,X^I+ w\,\bar X^I\right) \varepsilon^{ik} \mu_{I\,kj} =0
  \,, \nonumber\\
  R(\mathcal{V})_{23}^-{}^i{}_j = &\,
  R(\mathcal{V})_{23}^+{}^i{}_j =
  \tfrac12\,R(\mathcal{V})_{23}{}^i{}_j =
  -\tfrac12\mathrm{i}\,v_2^{-1}\,\sigma_3{}^i{}_j  \,,\nonumber\\
  D=&\, -\tfrac16\big(v_1^{-1} +2v_2^{-1}\big)\,,
  \nonumber\\
  v_1^{-1} =&\, \tfrac14 |w|^2 \,.
\end{align}

From the second line in \eqref{eq:generic}, combined with \eqref{eq:V}, we obtain the condition
\begin{equation}
 R(\mathcal{V})_{23}{}^i{}_j = 2\, \varepsilon^{ik}\mu_I{}_{kj}\, F_{23}^I = - \mathrm{i}\,v_2^{-1}\,\sigma_3{}^i{}_j\,,
\end{equation}
which by \eqref{eq:gauge-fix-generator} implies that
\begin{equation}\label{eq:dir-quant}
 t^{i}{}_{j} = \mathrm{i}\, \sigma^3{}^{i}_{j}\ , \qquad
 2\ \xi_I\ F_{23}^I = v_2^{-1} \,.
\end{equation}
The former condition simply identifies the spinor projection in \eqref{eq:KS-ansatz} with the generator controlling the abelian gauging of the compensating hypermultiplet. We remind the reader that the choice for $\sigma^3$ is conventional and that the projection may be defined using any $SU(2)$ generator. Our analysis then shows that this $SU(2)$ generator must be identified with the (also arbitrary) generator appearing in the gauging. The second of \eqref{eq:dir-quant} is a nontrivial condition on the field strengths that will turn out to correspond to a Dirac quantisation condition.

\subsection{Final set of off-shell equations}
\label{subsec:finalbps}
Summarizing, we are left with the following set of equations for the half-BPS near-horizon geometry of consideration,
\begin{empheq}[box=\fbox]{align}\label{eq:final}
\nonumber D &= - \frac16 (v_1^{-1} + 2 v_2^{-1})\ , \\
\nonumber \bar{w} \xi_I X^I &= - w \xi_J \bar{X}^J\ , \\
|w|^2 &= -8 {\rm i} w \xi_I \bar{X}^I = 4 v_1^{-1}\ , \\
\nonumber \xi_I p^I &= \frac{\kappa}{2}\ ,
\end{empheq}
where we used \eqref{eq:gauge-fix-generator} to derive the second and third relation from \eqref{eq:generic} and in the last identity we used \eqref{eq:dir-quant} and the definition for the magnetic charges,
\begin{align}
F^I_{23} = \frac{p^I}{v_2}\ .
\end{align}
Note the appearance of the quantity $\kappa$ in the last equation, which should just be equal to $1$ if we follow the above derivation. In fact $\kappa$ is related to the sign of the curvature of the horizon topology, and since so far we only looked at a spherical horizon we trivially find $\kappa_{S^2} = 1$. If we instead look at a higher genus Riemann surface, we find $\kappa_{T^2} = 0$, and $\kappa_{\Sigma_g^2} = -1$ for $g > 1$. This is the only way the BPS attractor equations change with the change of horizon topology, as already remarked in various papers \cite{Cacciatori:2009iz,Dall'Agata:2010gj,Katmadas:2014faa,Halmagyi:2014qza}.

The last equation we write is obtained by combining \eqref{eq:vectorBPS} and \eqref{eq:scalarBPS}
\begin{align}
\label{eq:attractorQ}
2\,{\rm i}\varepsilon_{ij}\hat F_{23}^{I\,-}\sigma_{3\;\;k}^{\;\;j} + Y^I_{ik} - 4\,X^I \bar X^J\,\mu_{J\,ik} = 0\ ,
\end{align}
from which we will define the magnetic charges, once we plug in the equation of motion for the auxiliary fields $Y_{ij}^I$.

The above equations are valid always for any choice of off-shell lagrangian, and need to be supplemented by the explicit equations of motion for the fields $D$ and $Y$, together with the Maxwell equations and Bianchi identities. Thus a complete solution can only be obtained after a choice of an explicit lagrangian, to which we proceed in the next section.

\section{$\mathcal{C}^2$ (F-term) action}
\label{sec:f-term}
The two derivative actions of vectors and hypers coupled to $N=2$ conformal supergravity are given in \cite{de Wit:1980tn,de Wit:1984px} and further generalized to four derivatives years after in \cite{deWit:2011gk,deWit:1999fp}. The full four-derivative F-term theory for vector and hypermultiplets coupled to conformal supergravity, with vanishing fermions, is given by:
\begin{align}\label{eq:4dtheory}
8\,\pi\,e^{-1}\cL_{\mathrm{vectors+hypers}}=&\,
 \mathrm{i} {\cal D}^{\mu} F_I \, {\cal D}_{\mu} \bar X^I - \mathrm{i} F_I\,\bar X^I
 (\ft16  \cR - D)
-\ft18\mathrm{i}  F_{IJ}\, Y^I_{ij} Y^{Jij}
\nonumber\\
&\,+\ft14 \mathrm{i} F_{IJ} (F^{-I}_{ab} -\ft 14 \bar X^I
T^-_{ab})(F^{-Jab} -\ft14 \bar X^J
T^{-\,ab})
\nonumber\\
&\,-\ft18 \mathrm{i} F_I(F^{+I}_{ab} -\ft14  X^I
T^+_{ab}) T^{+\,ab}-\ft1{32} \mathrm{i}\, F (T^+_{ab})^2
\nonumber\\
&\,- \ft14 \mathrm{i} \hat
B_{ij}\,F_{\hat{A}I}  Y^{Iij}   +\ft12 \mathrm{i} \hat G^{-ab}\, F_{\hat{A}I} (F^{-I}_{ab} - \ft14  \bar X^I
T^-_{ab})
\nonumber \\
&\,+\ft12 \mathrm{i} F_{\hat{A}}
\hat C -\ft18 \mathrm{i} F_{\hat{A}\hat{A}}(\varepsilon^{ik}
\varepsilon^{jl}  \hat B_{ij}
\hat B_{kl} -2 \hat G^-_{ab}\hat G^{-ab})
+ {\rm h.c.}
\nonumber \\
&\, - \ft12 \varepsilon^{ij}\, \bar \Omega_{\a\b} \,
{\cal D}_\mu A_i{}^\a \,{\cal D}^\mu  A_j{}^\b
+\chi (\ft16 \cR+  \ft12 D) \;,
\nonumber\\
&+2 G_{\bar\a\b}A^{i\bar\a}\,\xi_I \xi_J \bar X^I X^J ( t^\b{}_\g\, t^\g{}_\d) A_i{}^\d
\nonumber\\
&-\tfrac12 A_i{}^\a \bar\Omega_{\a\b}\,\xi_I Y^{I ij}t^\b{}_\g A_j{}^\g\ .
\end{align}
The components of the composite chiral multiplet made out of covariant quantities of the Weyl multiplet, denoted by hats, are still to be substituted to find the fully explicit result. We only require the modified homogeneity property
\begin{equation*}
X^I F_I + 2\,\hat A F_{A} = 2\,F\,.
\end{equation*}

We stress again that the way to recover the lagrangian of ungauged supergravity is by putting all FI parameters $\xi_I$ to zero.

We additionally need to solve the equations of motion for $Y^I_{i j}$ and $D$. First let us consider the e.o.m. of $Y$, which reads:
\begin{align}
\label{eq:EoM-Y}
 N_{I J} Y^J_{i j} - i \left( F_{A I} \hat{B}_{i j} - \varepsilon_{i l} \varepsilon_{j k} \hat{B}^{l k} \bar{F}_{A I} \right) - 2\,\chi_H \mu_{i j I} = 0\Rightarrow
 \nonumber\\
 Y_{ij}^I=2\,N^{IJ}\,\chi_H\,\mu_{i j I} + i\,N^{IJ}\left( \hat{B}_{i j} F_{A I} - \varepsilon_{i l} \varepsilon_{j k} \hat{B}^{l k} \bar{F}_{A I} \right)\ ,
\end{align}
where $N_{IJ}=-i(F_{IJ}-\bar F_{IJ})$ and $N^{IJ}$ is its inverse. This fixes $Y$ in terms of the gaugings in the two-derivative case when $F_{A I} = 0$.
On the other hand, for the same Lagrangian the $D$ equation of motion leads to the requirement
\begin{empheq}[box=\fbox]{align}
\label{eq:EoM-D}
&e^{-K} + \frac{1}{2} \chi_H + 192 i D (F_{\hat A}-\bar{F}_A) - 8 i (T^{- a b} F_{A I} \hat{F}^{- I}_{a b}-T^{+ a b} \bar{F}_{A I} \hat{F}^{+ I}_{a b})
\nonumber\\
 &- 8 i (T^{- a b} F_{A A} \hat{G}^{- }_{a b} - T^{+ a b} \bar{F}_{A A} \hat{G}^{+}_{a b}) = 0\ .
\end{empheq}

\subsection{Standard gauge fixing}
\label{subsec:feom}

If we concentrate on the two-derivative part of the lagrangian for the moment, we can perform the gauge fixing and match all our equations to the existing ones in the on-shell literature so that we can later compare and see how the higher derivatives change the results. From \eqref{eq:EoM-Y} in the two derivative case we find:
\begin{align}
 Y_{i j}^I = 2 N^{I J} \chi_H\,\mu_{i j\,J} \ ,
\end{align}
which would eventually lead to the usual on-shell supergravity scalar potential upon substitution in the lagrangian. The $D$ equation of motion imposes that
\begin{align}
 \chi_H = - 2\,e^{-K}\ ,
\end{align}
and choosing the gauge fixing condition
\begin{equation}\label{eq:gauge-fix-chi-2}
\chi_H = -2\ ,
\end{equation}
leads to the standard Poincare supergravity coupled to vector multiplets. The FI gauging can now be interpreted as nontrivial charges for the gravitino, once \eqref{eq:V} is used. Even if the higher derivative corrections change the e.o.m. for $Y$ we could consistently keep the same gauge fixing procedure and stick to \eqref{eq:gauge-fix-chi-2} if needed.

\subsection{Wald entropy}
\label{subsec:fwald}
To evaluate the entropy of a space-time configuration, one uses Wald's formula for a generic theory of gravity, which formally reads:
\begin{equation}
\label{eq:entropy_general}
S=\frac{1}{4} \int d^2\Omega E^{abcd}\,\varepsilon_{ab}\varepsilon_{cd}
\end{equation}
where, in our case, $d^2\Omega= v_2 \sin\theta\ {\mathrm d}\theta\ {\mathrm d}\phi$, $\varepsilon_{ab}$ is the binormal tensor, normalized to $\varepsilon_{ab}\,\varepsilon^{ab}=-2$ and has only $(0,1)$ components, perpendicular to the surface spanned by ${\mathrm d} \theta$ and ${\mathrm d} \phi$. So we can take locally $\varepsilon_{01}=-\varepsilon_{10}=1$. The normalization is in principle not fixed by the Noether procedure but can be easily derived by the requirement of finding the Bekenstein-Hawking entropy with the correct prefactor, see below. Finally, $E^{abcd}$ is the equation of motion for the Riemann tensor as if it were an independent field, and it reads:
\begin{align}
\label{eq:E_riemann}
E^{abcd}=&-\tfrac{\rm 1}6\Big(e^{-K}-\chi_H\Big)\eta^{a[c}\eta^{d]b}
\nonumber\\
&-8{\rm i}F_{AI}\,[\hat{F}^{-\,I\,cd}\,T^{-\,ab}
-2\eta^{\big[b[d}\eta^{c]e}\hat{F}^{-\,I}_{ef}\,T^{-\,a\big]f}
+\tfrac13\,\eta^{a[c}\eta^{d]b}\,\hat{F}^{-\,I}_{ef}\,T^{-\,ef}]+{\rm h.c.}
\nonumber\\
&+64{\rm i}\,F_A[\mathcal{R}(M)^{-\,abcd}-2\,\eta^{\big[b[d}\eta^{c]e}\mathcal{R}(M)^{-\,a\big]f}_{ef}+\tfrac13\,\eta^{a[c}\eta^{d]b}\,\mathcal{R}(M)^{-\,ef}_{ef}]+{\rm h.c.}
\nonumber\\
&+4{\rm i} F_A\,T^{-\,\big[ah}\,T^+_{gh}\,\eta^{b\big][d}\,\eta^{c]g}+{\rm h.c.}
\nonumber\\
&-8{\rm i}\,F_{AA}[\hat{G}^{-\,cd}\,T^{-\,ab}-2\eta^{\big[b[d}\eta^{c]e}\hat{G}^-_{ef}\,T^{-\,a\big]f}+\tfrac13\,\eta^{a[c}\eta^{d]b}\,\hat{G}^-_{ef}\,T^{-\,ef}]+{\rm h.c.}
\end{align}
Now, plugging in the values for the near-horizon field configuration we find
\begin{align}\label{eq:intermediate-entropy}
\cS=& \frac{1}{4} \int d^2\Omega\Big[
\tfrac13\,\big(e^{-K}-\chi_H \big)+
\nonumber\\
&\quad+\Big(\tfrac{64}3\,{\rm i}\,w\,F_{AI}\,\hat{F}^-_{01}{}^I-\tfrac{64}3{\rm i}\,F_A \, R
+8\,{\rm i}\,|w|^2\,F_A+\tfrac{1024}3\,{\rm i}\,w^2\,F_{AA}(D+\tfrac13 R)+{\rm h.c.}\Big)\Big]\,.
\end{align}
Note that the normalization we chose is in agreement with the Bekenstein-Hawking entropy formula. In fact, at the two derivative level the above formula simplifies completely since the second row vanishes and $\chi_{\mathrm H} = - 2 e^{-K}$ from the \eqref{eq:EoM-D}, such that
\begin{equation}
\cS= \frac{1}{4} \int\,e^{-K}\ v_2\,\sin\theta\, \mathrm{d}\theta\,\mathrm{d}\phi =\frac{\mathrm{A}}{4 G_4}\ ,
\end{equation}
since $e^{-K}$ is just the four-dimensional Newton constant in this case (c.f. \eqref{eq:4dtheory}). Observe that neither the horizon area nor the Newton constant are gauge invariant quantities in conformal supergravity by themselves, but the entropy is (as needed for a physical quantity). Thus gauge fixing is not needed here.
Using the equation of motion for $D$, we can further rewrite \eqref{eq:intermediate-entropy} as
\begin{align}\label{eq:trueentropy}
\cS=& \frac{1}{4} \int d^2\Omega\Big[
e^{-K}
-16\,\left( |w|^2 - \tfrac83 R + 16\,D \right)\,\mathrm{Im}F_A \Big]\,.
\end{align}
Note that the above formula holds for both the gauged and the ungauged solutions since we did not yet plug in any BPS conditions and gaugings terms do not enter. Now, upon using the expressions for the Ricci scalar and the BPS conditions in the gauged theory, \eqref{eq:final}, we obtain
\begin{align}
\cS= \frac{1}{4} \, \int d^2\Omega \Big[
 e^{-K} +64\,v_1^{-1}\,\mathrm{Im}F_A \Big]
\nonumber
\end{align}
\begin{equation}\label{eq:entropy_final}
\Rightarrow \boxed{ \cS =\pi e^{-K}\,v_2 +64\,\pi\,\frac{v_2}{v_1}\,\mathrm{Im}F_A \,.}
\end{equation}

Instead for the ungauged, fully BPS, attractor with $R^2$ interactions, one finds $R=D=0$ and $v_1 = v_2 = 16 |w|^{-2}$, so
that the entropy formula \eqref{eq:trueentropy} leads to
\begin{align}
\cS =&\, \pi e^{-K}\,v_2 - 256\,\pi\,\mathrm{Im}F_A \,.
\end{align}
This is not the limit $v_2=v_1$ of the result for the gauged case, as one might expect, since $D$
is negative definite for the gauged theory and cannot be continuously put to zero.

\subsection{Attractor equations}
\label{subsec:fmodels}
The attractor equations are given by \eqref{eq:attractorQ}, by using the equation of motion \eqref{eq:EoM-Y} for $Y_{ij}^I$, together with the near-horizon value for the field $\hat B_{ij}$ (see \eqref{eq:W-squared}). They read:
\begin{align}
\label{eq:F-_attract}
& \hat{F}^{-\,I}_{23}=-\Big[128\,\mathrm{i}\,v_2^{-1}\,N^{IJ}\big(X^K\,F_{AJ}-\bar X^K\,\bar F_{AJ}\big)+ \big(N^{IK}\,\chi_H-2\,X^I\bar X^K\big)\Big] \xi_{K}\,.
\end{align}

The physical vector fields are given by
\begin{align}
 F^{-\,I}_{23}=&\,
 \frac{\mathrm{i}}{4}\,w\,\bar{X}^I - \big(N^{IK}\,\chi_H-2\,X^I\bar X^K\big)\,\xi_{K}
+256\,v_2^{-1}\,N^{IJ}\mathrm{Im}\,\big(X^K\,\xi_{K}\,F_{AJ}\big)\,,
\end{align}
Let us obtain the dual vector fields, defined as
\begin{align}
 G^-_{ab\, I}&=(-2\,i)\frac{\partial \cL}{\partial F^{-\,I\,ab}}
=F_{IJ}\hat{F}^{-\,J}_{ab}+\tfrac14\,\bar F_I\,T^-_{ab}+\hat{G}^-_{ab}\,F_{AI}\,.
\end{align}
Using the BPS conditions \eqref{eq:final}, this becomes
\begin{align}
\label{eq:G-_attract}
G^-_{23\, I}
&=F_{IJ}\Big[256\,v_2^{-1}\,N^{JK}\mathrm{Im}(X^L\,F_{AK})-\big(N^{JL}\,\chi_H-2\,X^J\bar X^L\big)\Big]\xi_{L}
\nonumber\\
&\quad+\tfrac{i}4\,w\big(\bar F_I-64\,F_{AI}(D+\tfrac13\,R)\big)\,.
\end{align}

We can define electric and magnetic charges in a general way by integrating the real part of the gauge field strengths and their duals, as
\begin{equation}
 \Gamma \equiv \begin{pmatrix} p^I\\ q_I \end{pmatrix} = \int_{S^2} \begin{pmatrix} F^I\\ G_I \end{pmatrix}
 = v_2\,\begin{pmatrix} F^{-\,I}_{23} + F^{+\,I}_{23}\\ G^-_{23\, I} + G^+_{23\, I} \end{pmatrix}\,,
\end{equation}
which will be used in the following. The corresponding imaginary part is identified with the timelike components of the field strengths and reads
\begin{equation}
 \begin{pmatrix} F^{-\,I}_{23} - F^{+\,I}_{23}\\ G^-_{23\, I} - G^+_{23\, I} \end{pmatrix}
 =\mathrm{Re}\begin{pmatrix} \bar{w}\,X^I \\  \bar{w}\,F_I \end{pmatrix}
  + \chi_H\,\begin{pmatrix} 0 \\  \xi_I \end{pmatrix}\,,
\end{equation}
which is of exactly the same form as for the two derivative theory. Note however that the prepotential $F$ and $\chi_{\mathrm H}$ do carry information about the higher derivative corrections.

The attractor equations can be written using the definition for the charges and the explicit expressions for the gauge field strengths. The result is
\begin{empheq}[box=\fbox]{align}\label{eq:attractor}
 \Gamma\equiv \begin{pmatrix} p^I\\ q_I \end{pmatrix} = &\,
2 \Omega\cM(F)
\begin{pmatrix}
0 \\  v_2\,\chi_H\,\xi_J - 256\,\mathrm{Im}(X^K \xi_{K}\,F_{AJ})
\end{pmatrix}
 \nonumber\\
 &\,
+32\,v_2\,\mathrm{Im} \begin{pmatrix} 0\\ 2\,w\,F_{AI}(D+\tfrac13\,R) - \mathrm{i}\,w^2 F_{AI}\,\bar X^K \xi_{K}\end{pmatrix} \,,
\end{empheq}
where
\begin{align}
\label{eq:M(F)}
\cM (F)&=\begin{pmatrix}
\mathrm{Im}F +\mathrm{Re}F (\mathrm{Im}F)^{-1} \mathrm{Re}F &
-\mathrm{Re}F (\mathrm{Im}F)^{-1} \\
-\mathrm{Re}F (\mathrm{Im}F)^{-1} & (\mathrm{Im}F)^{-1}
\end{pmatrix}\ ,
\end{align}
using the matrix $F_{I J}$.

Using the above, we can compute a number of useful relations. Starting from the expressions for the field strengths, we can relate their inner product to the central charge, defined as
\begin{equation}
Z\equiv \mathrm{e}^{K/2}(F_I p^I - X^I q_I)\,.
\end{equation}
The result reads
\begin{equation}
 F_I F^{-\,I}_{23} - X^I G^-_{23\, I}
 = \frac{1}{2\,v_2}\,\mathrm{e}^{-K/2} Z + \frac14\,w\,\left( \mathrm{e}^{-K} + \frac14\,\chi_H \right)\,,
\end{equation}
and can be used to rewrite the equation of motion for the scalar field $D$, \eqref{eq:EoM-D}, as
\begin{equation}\label{eq:usefulidentity}
  2\ \mathrm{e}^{-K} \,v_2 =
 -\mathrm{Re}\left( \frac{Z}{\mathrm{i}\,\mathrm{e}^{K/2} \xi_I X^I} \right)
 -128\,\left( \frac{v_2}{v_1} + 2 \right)\,\mathrm{Im}F_A\,.
\end{equation}

Finally, by direct computation using the expression for the charges, one can find the following expression for the central charge
\begin{align}
\frac{Z}{\mathrm{i}\,\mathrm{e}^{K/2} \xi_I X^I} = &\,
\chi_H \,v_2 +256\,\mathrm{i}\,F_{A\,I}X^I
\nonumber\\
&\,
 - 1024\,\mathrm{i}\,F_{A\,I}\,N^{IJ}\left( v_2\,\chi_H\,\xi_J - 256\,\mathrm{Im}(X^K \xi_{K}\,F_{AJ})\right)\,X^L \xi_{L}
\,,
\end{align}
which can be rewritten as
\begin{align}
\frac{Z}{\mathrm{i}\,\mathrm{e}^{K/2}  \xi_I X^I} = &\,
\chi_H \,v_2
+256\,\mathrm{i}\,F_{A\,I}\left( X^I -\tfrac{\mathrm{i}}{4}\,w\,p^{I}\right)
\,,
\end{align}
upon use of the attractor equations.

If we now use the definition
\begin{equation}
    L \equiv \mathrm{e}^{K/2} \xi_I X^I\ ,
\end{equation}
as introduced in \cite{Dall'Agata:2010gj}, in the two-derivative case when $\chi_H = - 2\, \mathrm{e}^{-K}$ (and clearly $F_A = F_{A I} = 0$) we find
\begin{equation}
    \mathrm{Re} \left( \frac{Z}{i L} \right) = \left( \frac{Z}{i L} \right) = - \frac{2}{\pi} \cS\ .
\end{equation}
Now considering the more general higher derivative case and combining \eqref{eq:entropy_final} with \eqref{eq:usefulidentity} we instead find the following useful relation
\begin{equation}
\boxed{\cS = - \frac{\pi}{2} \mathrm{Im} \left( \frac{Z}{L} + 256\, F_A \right)\ .}
\end{equation}

\subsection{Relation to microscopic index}
\label{subsec:micro}
Note that there is another possibility for the gauge fixing condition, which is less standard in literature and differs from the one described in \ref{subsec:feom}. It was shown to be particularly useful when relating the gravity and the field theory side via the AdS/CFT correspondence, and amounts to choosing
\begin{equation}\label{eq:newgauge}
    \xi_I\, X^I = 1\ ,
\end{equation}
or any other arbitrary real constant without any further loss of generality. This means that $L = \mathrm{e}^{K/2}$ and then one looks at the much simpler expression
\begin{equation}\label{eq:R}
  {\cal R} \equiv \frac{Z}{L} = F_I p^I - X^I q_I\ .
\end{equation}
The upshot now is that the quantity ${\cal R}$ turns out to be functionally equivalent to the Witten index of the dual field theory, as explained in \cite{Benini:2015eyy}. It is therefore tempting to speculate that our higher derivative corrections to the relation between the Wald entropy and the quantity ${\cal R}$ provide the corresponding change in the Witten index of the dual field theory with finite $N$ corrections. We therefore define a new quantity
\begin{equation}
  {\cal R}_A \equiv F_I p^I - X^I q_I + 256\,F_A\ ,
\end{equation}
that knows about the higher derivative corrections to the entropy in the gauge \eqref{eq:newgauge},
\begin{equation}\label{eq:RA}
  \boxed{\cS = - \frac{\pi}{2}\,\mathrm{Im} {\cal R}_A \ .}
\end{equation}

\section{Examples with different asymptotics}
\label{sec:conclusion}
Here we consider several simple models to illustrate explicitly the formalism above. We choose three models with different asymptotics, which are well-known and understood in the two-derivative theory. Note that the addition of higher derivative terms in principle changes not only the near-horizon solution, but also the asymptotics. One then needs to be careful about the existence of the asymptotic space, which is sometimes guaranteed by supersymmetry.

However for the higher derivative lagrangian that we consider even non-supersymmetric vacua of the two-derivative theory are guaranteed to remain intact in the four-derivative theory. Asymptotically we can show that the cosmological constant does not change, for the following reason. The scalar potential in the lagrangian \eqref{eq:4dtheory} arises after substituting the auxiliary field $Y$ with its equation of motion, \eqref{eq:EoM-Y}. Notice there that the terms distinguishing the two-derivative and the four-derivative case depend linearly on the field $\hat{B}_{i j}$. From \eqref{eq:W-squared} we see that $\hat{B}_{i j}$ is proportional to the curvature of the $SU(2)$ gauge field ${\cal V}_{\mu}{}_i^{}\,{}^j$. On a maximally symmetric vacuum (and also depending on the case for other less symmetric vacua) vectors are constrained to vanish for symmetry reasons, and therefore one immediately finds the same value of scalar potential as in the two-derivative case since $\hat{B}_{i j} = 0$. Thus Minkowski$_4$ and AdS$_4$ remain asymptotic vacua automatically, and the same is true for many other vacua of interest such as the hvLif$_4$ that we also consider in one of the examples below\footnote{This is because even if hvLif$_4$ is not maximally symmetric, it is the reduction of the maximally symmetric AdS$_5$ \cite{Hristov:2014eza}.}.

\subsection{Minkowski}
Let us consider the so called T$^3$ prepotential
\begin{equation}
F=\frac{(X^1)^3}{X^0}+c\,\hat A\,\frac{X^1}{X^0}\ ,
\end{equation}
with the choice of FI parameters
\begin{equation}
 \xi_0 \neq 0\ , \qquad \xi_1 = 0\ .
\end{equation}
The two-derivative case is recovered when the parameter $c$ is taken to zero. In this case one finds a supersymmetric near-horizon geometry and a supersymmetry breaking flat space asymptotically, since the scalar potential identically vanishes. These black holes were shown to coincide with the more standard extremal non-BPS black holes in ungauged supergravity \cite{Hristov:2012nu}.

We will further make the simplifying assumption that we look at axion-free solutions, and adopt the notation
\begin{align}
&\mathrm{Re}\Big(\frac{X^1}{X^0}\Big)=0\;\Rightarrow \frac{X^1}{X^0}=i\,t \;\;,\;(t\in \mathbb{R})\ ,
\nonumber\\
&\frac{\hat{A}}{(X^0)^2}=\frac{-4\,w^2}{(X^0)^2}=-s\;\;,\; (s\in\mathbb{R})\ .
\end{align}

We first find from \eqref{eq:final}
\begin{align}
\sqrt{s}= i\,\sqrt{\tilde{s}}=2\,\frac{w}{X^0}=16\,i\,\xi_0\ ,
\end{align}
together with
\begin{align}
v_1^{-1} = 16 \xi_0^2 |X^0|^2\ .
\end{align}
These can be directly plugged in the attractor equations \eqref{eq:attractor} to find the following set of higher derivative equations:
\begin{align}
\label{eq:charges_case2}
p^1 &= q_0 = 0\ , \nonumber\\
\frac{p^0}{v_2}&=\frac{\xi_0}{v_2}\Big[\frac{\hat{c}\ t+v_2\,\chi_H}{2\,t\,(t^2 -\hat{c}\ \xi_0^2)}\Big]\ ,
\nonumber\\
\frac{q_1}{v_2}&=\frac{\xi_0}{v_2}\Big[\frac{(\hat{c}\,t+v_2\,\chi_H)(3\,t^2-\hat{c}\ \xi_0^2)}{2\,t\,(t^2-\hat{c}\ \xi_0^2)}\Big]
-\hat{c}\ \xi_0 v_2^{-1}\ ,
\end{align}
where we used $\hat{c} \equiv 256\ c$ for brevity. Note that $p^1=q_0=0$ just as in the two-derivative axion-free case \cite{Hristov:2012nu}.
It is now simple to derive the solution for the the physical scalar,
\begin{align}
    t = - \sqrt{\frac{q_1}{3 p^0}+ \frac{\hat{c}}{4 (p^0)^2}}\ ,
\end{align}
and for the radii
\begin{align}
    v_1^{-1} = 16 \xi_0^2 |X^0|^2\ , \quad v_2 = \frac{t}{\xi_0^2\,\chi_H}(t^2-2\,\hat{c}\,\xi_0^2) \,
\end{align}
where we used the condition $2 \xi_0 p^0 = 1$ (in the asymptotically Minkowski case we can only have spherical horizons, therefore no other choice of $\kappa$ is possible in \eqref{eq:final}).
The last equation we need to determine the complete solution is the equation of motion for the field $D$, \eqref{eq:EoM-D}. It is instructive first to look only at the two-derivative case when $\hat{c} = 0$. In this case we simply find
\begin{align}
    \chi_H = - 2 e^{-K} = 16 t^3 |X^0|^2\ ,
\end{align}
which immediately leads to
\begin{align}
    v_1^{-1} = 16 \xi_0^2 |X^0|^2\ = v_2^{-1}\ .
\end{align}
Note that the radii are not physical parameters as they depend on the arbitrary factor $|X^0|$. They are thus not gauge invariant, as we already remarked in section \ref{subsec:fwald}. Nevertheless, the above solution leads us to the gauge invariant conclusion that
\begin{align}
    \frac{v_2}{v_1} = 1\ ,
\end{align}
as expected from the precise agreement of these black hole attractors with the ones in ungauged supergravity \cite{Hristov:2012nu}. Now let us consider how the higher derivative corrections change the equation of motion for the field $D$ that ultimately tells us the relation between $\chi_H$ and $|X^0|^2$ and determines the ratio between the two radii. The general case of \eqref{eq:EoM-D} when $\hat{c} \neq 0$ leads to the following solution for $\chi_H$:
\begin{align}
    \chi_H = 16 \,\frac{|X^0|^2}{t}(t^2 - \hat{c}\,\xi_0^2)(t^2 - 2\,\hat{c}\,\xi_0^2)\ ,
\end{align}
leading to
\begin{align}
    \frac{v_2}{v_1} = \frac{t^2}{t^2 - \hat{c}\,\xi_0^2} = 1 + \frac{3\,\hat{c}}{4\,p^0\,q_1}
\end{align}
It is now easy to see the difference between gauged ($\xi_0 \neq 0$) and ungauged ($\xi_0 = 0$) supergravity. In the two derivative case both lead to the same attractor solution, while the higher deivative case when $\hat{c} \neq 0$ we get diverging results. In ungauged supergravity the ratio $v_1/v_2$ must always be $1$ as the Ricci scalar must vanish. This is no longer the case in gauged supergravity, even though asymptotically we still have a Minkowski vacuum. This suggests that from a full quantum gravity point of view the two-derivative equivalence is only a coincidence. Note however that in a more restricted setting where higher order corrections are directly dictated by string theory one might again find the same equivalence, possibly on a case by case basis.

Finally, for completeness we give the full entropy in terms of the electromagnetic charges, which now reads
\begin{align}
 \cS = \frac{2\,\sqrt{3}\pi}{9\,p^0}\left(q_1\,p^0 + \frac{3}{4} \hat{c}\right)^{3/2}\ .
\end{align}

\subsection{hvLif}
Our next example makes use of the same prepotential
\begin{equation}
F=\frac{(X^1)^3}{X^0}+c\,\hat A\,\frac{X^1}{X^0}\ ,
\end{equation}
but an orthogonal choice of FI parameters
\begin{equation}
 \xi_0 = 0\ , \qquad \xi_1 \neq 0\ .
\end{equation}
In the two-derivative case of $c=0$ the black holes solutions were analyzed carefully \cite{Hristov:2014eza} and were shown to originate from dimensional reduction of AdS$_5$ black strings. The 4d solutions therefore exhibit a runaway behavior, with an asymptotic solution called hyperscaling-violating Lifshitz, or hvLif. The full solution in this case is quarter-supersymmetric. Note that in the STU prepotential the asymptotically hvLif black holes can have horizon topologies of any Riemann surface, but for the simplified T$^3$ prepotential considered above there are only hyperbolic (i.e.\ higher genus) solutions at a two-derivative level. However as we will soon see the higher derivative terms change the situation and so we consider $\xi_I p^I = \xi_1 p^1 = \kappa/2$ for any $\kappa = \{-1, 0, 1 \}$.

Considering the $c \neq 0$ case, we again take the axion-free assumption, and use the parametrizations
\begin{align}
&\mathrm{Re}\Big(\frac{X^1}{X^0}\Big)=0\;\Rightarrow \frac{X^1}{X^0}=i\,t \;\;,\;(t\in \mathbb{R})\ ,
\nonumber\\
&\frac{\hat{A}}{(X^0)^2}=\frac{-4\,w^2}{(X^0)^2}=-s\;\;,\; (s\in\mathbb{R})\ .
\end{align}

We then find from \eqref{eq:final}
\begin{align}
\sqrt{s}=2\,\frac{w}{X^0}= -16\,t\,\xi_1\ .
\end{align}
together with
\begin{align}
v_1^{-1} = 16 t^2 \xi_1^2 |X^0|^2\ .
\end{align}

Here we have the same prepotential as before which leads to the attractor equations \eqref{eq:attractor}
\begin{align}
\label{eq:charges_case1}
p^0&= q_1 = 0\ ,
\nonumber\\
\frac{p^1}{v_2}&=\frac{\xi_1}{6\,v_2\,t}\left( \hat{c}\,t-\chi_H\,v_2\right)\ ,
\nonumber\\
\frac{q_0}{v_2}&=\frac{\xi_1\,t}{6\,v_2}\left(\hat{c} \, t (-3 +  \hat{c}\, \xi_1^2) - v_2\,\chi_H \, (3 + \hat{c}\, \xi_1^2) \right)\ ,
\end{align}
where we again used $\hat{c} \equiv 256 c$. We see that $p^0=q_1=0$ as expected from the two-derivative axionless case \cite{Hristov:2014eza}. Further using the equation of motion for the $D$ field, \eqref{eq:EoM-D}, we eventually find
\begin{align}
t = -\frac{\sqrt{q_0}}{\sqrt{- \hat{c} \xi_1 + p^1 (3 + \hat{c} \xi_1^2)}}\ ,\qquad v_2=\frac{2\,\hat{c}\,\xi_1^2 + 2\,\xi_1\,p^1 (-3 +\hat{c}\,\xi_1^2)}{16\,t^2\,\xi_1^2\,|X^0|^2}\ ,
\end{align}
and therefore
\begin{align}
\frac{v_2}{v_1} = 2\,\hat{c}\,\xi_1^2 + \kappa (-3 +\hat{c}\,\xi_1^2) \ ,
\end{align}
which reproduces the two-derivative result that $v_2 = 3 v_1$ for $\kappa=-1$. In the two-derivative limit $\hat{c}=0$ it is clear that any other choice for $\kappa$ is inconsistent, but this is no longer the case if $\hat{c} \neq 0$ is in some favorable parameter range. For $\kappa = 0$, i.e.\ toroidal horizon, it is enough that $\hat{c} > 0$ and we find the so called small black holes, with a vanishing classical horizon that appears after higher derivative corrections. For the spherical horizon with $\kappa = 1$ instead we find the condition for the appearance of horizon is $\hat{c} > \xi_1^{-2}$, which is more restrictive and subject to possible change from even higher derivative corrections, but still a valid possibility.

Let us finally see how the entropy gets corrected in the higher genus case $\kappa = -1$ (in the toroidal and spherical cases the entropy is proportional to $\hat{c}$ as it has no classical contribution),
\begin{align}
    S = 2 \sqrt{3}\,\pi \sqrt{q_0 p^1}\,\sqrt{ (p^1)^2 + \tfrac14\,\hat{c}}\ .
\end{align}
It would be interesting to see how this formula compares with higher derivative corrections to horizons in five-dimensional supergravity \cite{Baggio:2014hua} and then relate the microscopic descriptions \cite{Benini:2013cda,Hristov:2014eza}. We leave this for the future as one should in principle first repeat our exercise for all possible four-derivative terms in 4d gauged supergravity.

\subsection{AdS}
In the last example we consider asymptotically AdS$_4$ models, the simplest of which has the prepotential
\begin{equation}
F=-2\,{\rm i}\sqrt{X_0\,X_1^3+c\,\hat A\,X_0\,X_1}\sim -2\,{\rm i}\sqrt{X_0\,X_1^3}-{\rm i}\,c\,\hat A\,\sqrt{\frac{X_0}{X_1}}\ ,
\end{equation}
with two non-vanishing FI parameters which we already choose equal for simplicity
\begin{equation}
 \xi_0 = \xi_1 = \frac12\ ,
\end{equation}
so that the charge quantization condition is simply $p^0 + p^1 = \kappa$. In the two derivative case ($c=0$) this is a truncation of the STU model that arises from reduction of 11 dimensional supergravity on S$^7$. The full black hole solutions in such theories were first found in \cite{Cacciatori:2009iz} and further worked out in \cite{Dall'Agata:2010gj,Hristov:2010ri,Klemm:2011xw,Halmagyi:2013sla,Halmagyi:2013qoa,Katmadas:2014faa,Halmagyi:2014qza}. Their dual microscopic description was uncovered in \cite{Benini:2015eyy} with the successful match to the Bekenstein-Hawking entropy. It is therefore interesting to look at higher derivative corrections to such solutions which might lead to new tests of AdS/CFT and further insight into quantum gravity. Note however that the particular higher derivative corrections do not immediately follow from 11d supergravity, so the model is purely exemplary. We hope to come back to more direct applications of higher derivative corrections from string theory in the future.

Focusing on the explicit model, we again look at the case of axion-free solutions, which means
\begin{equation}
\frac{X^1}{X^0}=t^2 \; ,
\end{equation}
for a real scalar $t$. As before we also have
\begin{equation}
\frac{\hat{A}}{(X^0)^2}=\frac{-4\,w^2}{(X^0)^2}=-s\ ,
\end{equation}
leading to
\begin{align}
N_{IJ}=\begin{pmatrix} -t(t^2+\tfrac32\,c\,s\,) & \frac{c\,s}{2\,t}+3\,t \\
\frac{c\,s}{2\,t}+3\,t & \frac{c\,s + 6\,t^2}{2\,t^3}
\end{pmatrix}\ .
\end{align}
We now find from \eqref{eq:final}
\begin{align}
\sqrt{s}=8\,{\rm i}\,(1+t^2)\ ,
\end{align}
together with
\begin{align}
v_1^{-1} = 8\,(1+t^2)^2\,|X^0|^2\ .
\end{align}
The equation of motion for $D$, \eqref{eq:EoM-D}, now gives the relation
\begin{equation}
\chi_H + \frac{\hat c\,t}{v_2} = 2\,t |X^0|^2\big[8\,t^2-\hat c (1+t^2)^2\big)\big]\ .
\end{equation}
Now we first focus on the two-derivative case, in which the attractor equations \eqref{eq:attractor} are simply given by
\begin{align}
q_0&=q_1=0\ ,
\nonumber\\
\frac{p^0}{v_2}&=\frac{\chi_H (1-t^2)}{4\,t^3}\ ,
\nonumber\\
\frac{p^1}{v_2}&=-\frac{\chi_H (3+t^2)}{12\,t}\ ,
\end{align}
with the solution
\begin{equation}\label{eq:tsol}
t=-\sqrt{\frac{3(\kappa-2\,p^0)-\sqrt{3\,(\kappa-4\,p^0)\,(3\,\kappa-4\,p^0)}}{2\,p^0}}\ ,
\end{equation}
and
\begin{equation}
\label{eq:v_2sol}
v_2=\frac{4\,p^0\,t^3}{\chi_H\,(1-t^2)} = \frac{p^0}{4\,(1-t^2)\,|X^0|^2}\ ,
\end{equation}
after plugging in the condition $p^0+p^1 = \kappa$. We then find
\begin{equation}
  \frac{v_2}{v_1} = p^0\, \frac{(1+t^2)^2}{(1-t^2)}\ ,
\end{equation}
which we require for consistency to be a positive number. This sets the required ranges for the magnetic charge $p^0$ for the different choices of horizon topology, $\kappa=\{-1, 0, 1 \}$. Finally the entropy is given by
\begin{equation}
{\cal S}=\frac{2\,\pi\,p^0\,(-t)^3}{(1-t^2)}\ ,
\end{equation}
in terms of the scalar $t$ given above, \eqref{eq:tsol}.

The higher derivative equations in this case unfortunately look more complicated than in the previous examples,
\begin{align}
q_0&=q_1=0,
\nonumber\\
p^0&=-2\,\frac{v_2\,\chi_H(1-t^2)+\hat c\,t\,(1+t^2)}{t\,(-8\,t^2+\hat c\,(1+t^2)^2)}\ ,
\nonumber\\
p^1&=\frac{2t\,\Big[-\hat c\,t\,(1+t^2)\Big(8\,t^2+\hat c\,(1+t^2)^2\Big)+v_2\,\chi_H \Big(-8\,t^2\,(3+t^2)+\hat c (1+t^2)^2(1+3\,t^2)\Big)\Big]}{\Big(-24\,t^2+\hat c\,(1+t^2)^2\Big)\Big(-8\,t^2+\hat c\,(1+t^2)^2\Big)}\ .
\end{align}
By working to first order in $\hat{c}$ in all the equations, one finds the  modification to the two derivative near-horizon values $t = t_* + \hat{c}\, t' + {\cal O} (\hat{c}^2),\ v_2 = v_2^* + \hat{c}\, v_2' + {\cal O} (\hat{c}^2)$,
\begin{align}
t'&=\frac{3\,(1+t_*^2)+8\,|X^0|^2\,t_*^2\,v_2^*\,(1-t_*^4)}{48\,|X^0|^2\,t_*\,v_2^*\,(3-t_*^2)}\ ,
\nonumber\\
v_2'&=\frac{4\,v_2^*\,|X^0|^2\,t_*^2(1+t_*^2)-3}{12\,|X^0|^2\,(3-t_*^2)}\ ,
\end{align}
with $t_*$ and $v_2^*=\frac{p^0}{4(1-t_*^2)|X^0|^2}$ the corresponding two derivative solutions \eqref{eq:tsol} and \eqref{eq:v_2sol}, respectively. Eventually, the first order corrected entropy reads
\begin{align}
{\cal S}&=\frac{2\,\pi\,p^0\,(-t_*)^3}{(1-t_*^2)}-\pi\,\hat c\,\Big(t_*\,v_2^*(1-\tfrac23\,t_*^2-\tfrac53\,t_*^3+24\,t_*\,t')+8\,t_*^3\,\,v_2'\Big)\,|X^0|^2\ ,
\end{align}
in terms of the original two-derivative solution $t_*$ and $v_2^*$, and independent by inspection from the unphysical $|X^0|^2$ as expected. We spare the reader the explicit expressions for the other quantities as they are no particularly illuminating, making only the point that the higher order equations are always solvable in principle to any precision needed.

\section*{Acknowledgements}
We would like to thank Marco Baggio and Valentin Reys for useful discussions. IL would like to thank the University of Milano-Bicocca for hospitatility in the initial stage of this project. The work of SK is supported in part by INFN and by the ERC Grant 307286 (XD-STRING).  The work of IL is partly supported by a DST Ramanujan Grant.

\appendix

\section{Superconformal calculus}
\renewcommand{\theequation}{B.\arabic{equation}}
\label{B}
Throughout this paper, space-time indices are denoted by $\mu,\nu,\ldots$ and Lorentz
indices are denoted by $a,b,\ldots$. The metric signature is $\eta_{ab} = \textrm{diag}(-1,1,1,1)$
and we use Pauli-K\"all\'{e}n conventions, so that the anti-symmetric symbol
$\veps_{abcd}$ is imaginary, with $\veps_{0123} = -\ri$. We therefore define (anti-)selfdual
tensors as complex conjugates of each other, by
\begin{gather}
F_{ab}^\pm = \tfrac{1}{2} (F_{ab} \pm \tilde F_{ab})~, \qquad
\tilde F_{ab} = \tfrac{1}{2} \veps_{abcd} F^{cd}~, \qquad
\tilde F_{ab}^\pm = \pm F_{ab}^\pm~. \label{eq:Fselfdual}
\end{gather}
$\mathrm{SU}(2)$ indices are denoted by $i,j,\ldots$ and are raised and lowered by
complex conjugation, $(T_{abij})^* = T_{ab}{}^{ij}$, while the invariant $\rm SU(2)$ tensor
$\veps^{ij}$ and $\veps_{ij}$ is defined as $\veps^{12} = 1$ with $\veps^{ij} \veps_{kj} = \delta^i_k$.

The superconformal algebra includes the generators of the general coordinate, local Lorentz,
dilatation, special conformal, chiral $\mathrm{SU}(2)$ and
$\mathrm{U}(1)$, supersymmetry (Q) and special supersymmetry (S)
transformations. The gauge fields associated with general coordinate
transformations ($e_\mu{}^a$), dilatations ($b_\mu$), R-symmetry
($\mathcal{V}_\mu{}^i{}_j$ and $A_\mu$) and Q-supersymmetry
($\psi_\mu{}^i$) are independent fundamental fields.  The remaining gauge fields
associated with the Lorentz ($\omega_\mu{}^{ab}$), special conformal
($f_\mu{}^a$) and S-supersymmetry transformations ($\phi_\mu{}^i$) are
composite objects.
The multiplet also contains three matter fields: a Majorana spinor doublet
$\chi^i$, a scalar $D$, and a selfdual Lorentz tensor $T_{ab\,ij}$,
which is anti-symmetric in $[ab]$ and $[ij]$. The Weyl and chiral
weights have been collected in table \ref{table:weyl}.

Under Q-supersymmetry, S-supersymmetry and special conformal
transformations the Weyl multiplet fields transform
as
\begin{eqnarray}
  \label{eq:weyl-multiplet}
  \delta e_\mu{}^a & =& \bar{\epsilon}^i \, \gamma^a \psi_{ \mu i} +
  \bar{\epsilon}_i \, \gamma^a \psi_{ \mu}{}^i \, , \nonumber\\
  \delta \psi_{\mu}{}^{i} & =& 2 \,\mathcal{D}_\mu \epsilon^i - \tfrac{1}{8}
  T_{ab}{}^{ij} \gamma^{ab}\gamma_\mu \epsilon_j - \gamma_\mu \eta^i
  \, \nonumber \\
  \delta b_\mu & =& \tfrac{1}{2} \bar{\epsilon}^i \phi_{\mu i} -
  \tfrac{3}{4} \bar{\epsilon}^i \gamma_\mu \chi_i - \tfrac{1}{2}
  \bar{\eta}^i \psi_{\mu i} + \mbox{h.c.} + \Lambda^a_{\rm K} e_{\mu a} \, ,
  \nonumber \\
  \delta A_{\mu} & =& \tfrac{1}{2} \mathrm{i} \bar{\epsilon}^i \phi_{\mu i} +
  \tfrac{3}{4} \mathrm{i} \bar{\epsilon}^i \gamma_\mu \, \chi_i +
  \tfrac{1}{2} \mathrm{i}
  \bar{\eta}^i \psi_{\mu i} + \mbox{h.c.} \, , \nonumber\\
  \delta \mathcal{V}_\mu{}^{i}{}_j &=& 2\, \bar{\epsilon}_j
  \phi_\mu{}^i - 3
  \bar{\epsilon}_j \gamma_\mu \, \chi^i + 2 \bar{\eta}_j \, \psi_{\mu}{}^i
  - (\mbox{h.c. ; traceless}) \, , \nonumber \\
  \delta T_{ab}{}^{ij} &=& 8 \,\bar{\epsilon}^{[i} R(Q)_{ab}{}^{j]} \,
  , \nonumber \\
  \delta \chi^i & =& - \tfrac{1}{12} \gamma^{ab} \, \Slash{D} T_{ab}{}^{ij}
  \, \epsilon_j + \tfrac{1}{6} R(\mathcal{V})_{\mu\nu}{}^i{}_j
  \gamma^{\mu\nu} \epsilon^j -
  \tfrac{1}{3} \mathrm{i} R_{\mu\nu}(A) \gamma^{\mu\nu} \epsilon^i + D
  \epsilon^i +
  \tfrac{1}{12} \gamma_{ab} T^{ab ij} \eta_j \, , \nonumber \\
  \delta D & =& \bar{\epsilon}^i \,  \Slash{D} \chi_i +
  \bar{\epsilon}_i \,\Slash{D}\chi^i \, .
\end{eqnarray}
Here $\epsilon^i$ and $\epsilon_i$ denote the spinorial parameters of
Q-supersymmetry, $\eta^i$ and $\eta_i$ those of S-supersymmetry, and
$\Lambda_{\rm K}{}^a$ is the transformation parameter for special conformal
boosts.  The full superconformally covariant derivative is denoted by
$D_\mu$, while $\mathcal{D}_\mu$ denotes a covariant derivative with
respect to Lorentz, dilatation, chiral $\mathrm{U}(1)$ and
$\mathrm{SU}(2)$ transformations,
\begin{equation}
  \label{eq:D-epslon}
  \mathcal{D}_{\mu} \epsilon^i = \big(\partial_\mu - \tfrac{1}{4}
    \omega_\mu{}^{cd} \, \gamma_{cd} + \tfrac1{2} \, b_\mu +
    \tfrac{1}{2}\mathrm{i} \, A_\mu  \big) \epsilon^i + \tfrac1{2} \,
  \mathcal{V}_{\mu}{}^i{}_j \, \epsilon^j  \,.
\end{equation}
\begin{table}[t]
\begin{tabular*}{\textwidth}{@{\extracolsep{\fill}}
    |c||cccccccc|ccc||ccc| }
\hline
\noalign{\smallskip}
 & &\multicolumn{9}{c}{Weyl multiplet} & &
 \multicolumn{2}{c}{parameters} & \\[1mm]  \hline \hline
 \noalign{\smallskip}
 field & $e_\mu{}^{a}$ & $\psi_\mu{}^i$ & $b_\mu$ & $A_\mu$ &
 $\mathcal{V}_\mu{}^i{}_j$ & $T_{ab}{}^{ij} $ &
 $ \chi^i $ & $D$ & $\omega_\mu^{ab}$ & $f_\mu{}^a$ & $\phi_\mu{}^i$ &
 $\epsilon^i$ & $\eta^i$
 & \\[1mm] \hline\noalign{\smallskip}
$w$  & $-1$ & $-\tfrac12 $ & 0 &  0 & 0 & 1 & $\tfrac{3}{2}$ & 2 & 0 &
1 & $\tfrac12 $ & $ -\tfrac12 $  & $ \tfrac12  $ & \\[1mm] \hline\noalign{\smallskip}
$c$  & $0$ & $-\tfrac12 $ & 0 &  0 & 0 & $-1$ & $-\tfrac{1}{2}$ & 0 &
0 & 0 & $-\tfrac12 $ & $ -\tfrac12 $  & $ -\tfrac12  $ & \\[1mm] \hline\noalign{\smallskip}
 $\gamma_5$   &  & $\;+$ &   &    &   &   & $\;+$ &  &  &  & $\;-$ &
 $ \;+ $  & $\;-$ & \\ \hline
\end{tabular*}
\vskip 2mm
\renewcommand{\baselinestretch}{1}
\parbox[c]{\textwidth}{\caption{\label{table:weyl}{\footnotesize
Weyl and chiral weights ($w$ and $c$) and fermion
chirality $(\gamma_5)$ of the Weyl multiplet component fields and the
supersymmetry transformation parameters.}}}
\end{table}
Only a subset of the various covariant curvatures appears explicitly
in this work, given by
\begin{align}
  \label{eq:curvatures}
  R(P)_{\mu \nu}{}^a  = & \, 2 \, \partial_{[\mu} \, e_{\nu]}{}^a + 2 \,
  b_{[\mu} \, e_{\nu]}{}^a -2 \, \omega_{[\mu}{}^{ab} \, e_{\nu]b} -
  \tfrac1{2} ( \bar\psi_{[\mu}{}^i \gamma^a \psi_{\nu]i} +
  \mbox{h.c.} ) \, , \nonumber\\[.2ex]
  R(Q)_{\mu \nu}{}^i = & \, 2 \, \mathcal{D}_{[\mu} \psi_{\nu]}{}^i -
  \gamma_{[\mu}   \phi_{\nu]}{}^i - \tfrac{1}{8} \, T^{abij} \,
  \gamma_{ab} \, \gamma_{[\mu} \psi_{\nu]j} \, , \nonumber\\[.2ex]
  R(A)_{\mu \nu} = & \, 2 \, \partial_{[\mu} A_{\nu ]} - \mathrm{i}
  \left( \tfrac12
    \bar{\psi}_{[\mu}{}^i \phi_{\nu]i} + \tfrac{3}{4} \bar{\psi}_{[\mu}{}^i
    \gamma_{\nu ]} \chi_i - \mbox{h.c.} \right) \, , \nonumber\\[.2ex]
  R(\mathcal{V})_{\mu \nu}{}^i{}_j =& \, 2\, \partial_{[\mu}
  \mathcal{V}_{\nu]}{}^i{}_j +
  \mathcal{V}_{[\mu}{}^i{}_k \, \mathcal{V}_{\nu]}{}^k{}_j  +  2 (
    \bar{\psi}_{[\mu}{}^i \, \phi_{\nu]j} - \bar{\psi}_{[\mu j} \,
    \phi_{\nu]}{}^i )
  -3 ( \bar{\psi}_{[\mu}{}^i \gamma_{\nu]} \chi_j -
    \bar{\psi}_{[\mu j} \gamma_{\nu]} \chi^i ) \nonumber\\
& \, - \delta_j{}^i ( \bar{\psi}_{[\mu}{}^k \, \phi_{\nu]k} -
  \bar{\psi}_{[\mu k} \, \phi_{\nu]}{}^k )
  + \tfrac{3}{2}\delta_j{}^i (\bar{\psi}_{[\mu}{}^k \gamma_{\nu]}
  \chi_k - \bar{\psi}_{[\mu k} \gamma_{\nu]} \chi^k)  \, , \nonumber\\[.2ex]
  R(M)_{\mu \nu}{}^{ab} = & \,
  \, 2 \,\partial_{[\mu} \omega_{\nu]}{}^{ab} - 2\, \omega_{[\mu}{}^{ac}
  \omega_{\nu]c}{}^b
  - 4 f_{[\mu}{}^{[a} e_{\nu]}{}^{b]}
  + \tfrac12 (\bar{\psi}_{[\mu}{}^i \, \gamma^{ab} \,
  \phi_{\nu]i} + \mbox{h.c.} ) \nonumber\\
& \, + ( \tfrac14 \bar{\psi}_{\mu}{}^i   \,
  \psi_{\nu}{}^j  \, T^{ab}{}_{ij}
  - \tfrac{3}{4} \bar{\psi}_{[\mu}{}^i \, \gamma_{\nu]} \, \gamma^{ab}
  \chi_i
  - \bar{\psi}_{[\mu}{}^i \, \gamma_{\nu]} \,R(Q)^{ab}{}_i
  + \mbox{h.c.} ) \, .
\end{align}
The connections $\omega_{\mu}{}^{ab}$, $\phi_\mu{}^i$ and $f_{\mu}{}^a$
are algebraically determined by imposing the conventional constraints
\begin{gather}
  R(P)_{\mu \nu}{}^a =  0 ~, \qquad
  \gamma^\mu R(Q)_{\mu \nu}{}^i + \tfrac32 \gamma_{\nu}
  \chi^i = 0 ~, \nonumber \\
  e^{\nu}{}_b \,R(M)_{\mu \nu a}{}^b - \mathrm{i} \tilde{R}(A)_{\mu a} +
  \tfrac1{8} T_{abij} T_\mu{}^{bij} -\tfrac{3}{2} D \,e_{\mu a} = 0
  \,,  \label{eq:conv-constraints}
\end{gather}
which can be solved to obtain
\begin{align}
  \label{eq:dependent}
 \omega_\mu{}^{ab} =&\, -2e^{\nu[a}\partial_{[\mu}e_{\nu]}{}^{b]}
     -e^{\nu[a}e^{b]\sigma}e_{\mu c}\partial_\sigma e_\nu{}^c
     -2e_\mu{}^{[a}e^{b]\nu}b_\nu   \nonumber\\
      &\, -\ft{1}{4}(2\bar{\psi}_\mu^i\gamma^{[a}\psi_i^{b]}
     +\bar{\psi}^{ai}\gamma_\mu\psi^b_i+{\rm h.c.}) \,,\nonumber\\
 \phi_\mu{}^i  =& \, \tfrac12 \left( \gamma^{\rho \sigma} \gamma_\mu -
    \tfrac{1}{3} \gamma_\mu \gamma^{\rho \sigma} \right) \left(
    \mathcal{D}_\rho
    \psi_\sigma{}^i - \tfrac{1}{16} T^{abij} \gamma_{ab} \gamma_\rho
    \psi_{\sigma j} + \tfrac{1}{4} \gamma_{\rho \sigma} \chi^i \right)
    \,,  \nonumber\\
 f_\mu{}^a=&\, \tfrac12 R(\omega,e)_\mu{}^a -\tfrac14 \big(D+\tfrac13
   R(\omega,e)\big) e_\mu{}^a -\tfrac12\mathrm{i}\tilde R(A)_\mu{}^a +
   \tfrac1{16} T_{\mu b} {}^{ij} T^{ab}{}_{ij} +
    \mbox{fermions} \, .
\end{align}
We omitted the fermionic terms in $f_\mu{}^a$ for brevity and
$R(\omega,e)_\mu{}^a= R(\omega)_{\mu\nu}{}^{ab} e_b{}^\nu$ is
the non-symmetric Ricci tensor, with $R(\omega,e)$ the corresponding
Ricci scalar. The curvature $R(\omega)_{\mu\nu}{}^{ab}$ is associated
with the spin connection field $\omega_\mu{}^{ab}$.

The covariant objects of the Weyl multiplet form a so called reduced chiral multiplet,
whose components read:
\begin{align}
\label{cov-weyl-mult}
A_{ab} &=T_{ab}{}^{ij}\,\varepsilon_{ij}\,,
\nonumber\\
\psi_{abi} &=8\,\varepsilon_{ij} R(Q)_{ab}^j\,,
\nonumber\\
B_{abij} &=-8\,\varepsilon_{k(i}R(\cV)_{ab\;\;j)}^{-\;\,k}\,,
\nonumber\\
G^-_{ab}{}^{cd} &=-8\,\cR(M)_{ab}^{-\,cd}
\nonumber\\
\Lambda_{abi} &=8\,R(S)^-_{abi}  +6\,\gamma_{ab}\Slash{D}\chi_i
\nonumber\\
C_{ab} &=4\,D_{[a}\,D^c T_{b]cij}\,\varepsilon^{ij}-\mathrm{dual}\;.
\end{align}
We use a modified version of the Weyl tensor, $\cR(M)_{ab}^{-\,cd}$, defined as
\begin{align}
  \label{eq:mod-curv}
  \cR(M)_{ab}{}^{\!cd} =\, &    R(M)_{ab}\,^{cd} + \ft1{16}\big(
  T_{abij}\,T^{cdij} + T_{ab}{}^{ij}\, T^{cd}{}_{ij}  \big)\,,
\end{align}
while $R(S)_{ab}{}^i$ stands for the supercovariant curvature of the composite S-supersymmetry
gauge field, $\phi_\mu{}^i$, and its explicit expression is not important for this work.
By squaring the covariant Weyl multiplet in \eqref{cov-weyl-mult}, one obtains a scalar
chiral multiplet of weight $w=2$, with components
\begin{align}
  \label{eq:W-squared}
  \hat A   =&\,(T_{ab}{}^{ij}\varepsilon_{ij})^2\,,\nonumber \\[.2ex]
  \hat \Psi_i  =&\, 16\, \varepsilon_{ij}R(Q)^j_{ab} \,T^{klab}
  \, \varepsilon_{kl} \,,\nonumber\\[.2ex]
  \hat B_{ij}   =&\, -16 \,\varepsilon_{k(i}R({\cal
    V})^k{}_{j)ab} \, T^{lmab}\,\varepsilon_{lm} -64
  \,\varepsilon_{ik}\varepsilon_{jl}\,\bar R(Q)_{ab}{}^k\, R(Q)^{l\,ab}
  \,,\nonumber\\[.2ex]
  \hat G^{-ab}   =&\, -16 \,\cR(M)_{cd}{}^{\!ab} \,
  T^{klcd}\,\varepsilon_{kl}  -16 \,\varepsilon_{ij}\, \bar
  R(Q)^i_{cd}  \gamma^{ab} R(Q)^{cd\,j}  \,,\nonumber\\[.2ex]
 \hat \Lambda_i  =&\, 32\, \varepsilon_{ij} \,\gamma^{ab} R(Q)_{cd}^j\,
  \cR(M)^{cd}{}_{\!ab}
  +16\,({R}(S)_{ab\,i} +3 \gamma_{[a} D_{b]}  \chi_i) \,
  T^{klab}\, \varepsilon_{kl} \nonumber\\
  &\, -64\, R({\cal V})_{ab}{}^{\!k}{}_i \,\varepsilon_{kl}\,R(Q)^{ab\,l}
  \,,\nonumber\\[.2ex]
  \hat C  =&\,  64\, \cR(M)^{-cd}{}_{\!ab}\,
 \cR(M)^-_{cd}{}^{\!ab}  + 32\, R({\cal V})^{-ab\,k}{}_l^{~} \,
  R({\cal V})^-_{ab}{}^{\!l}{}_k  \nonumber \\
  &\, - 32\, T^{ab\,ij} \, D_a \,D^cT_{cb\,ij} +
  128\,\bar{R}(S)^{ab}{}_i  \,R(Q)_{ab}{}^i  +384 \,\bar
  R(Q)^{ab\,i} \gamma_aD_b\chi_i   \,.
\end{align}
These components appear in the lagrangian describing $R^2$ interactions,
arising both by the explicit $R^2$ term in $\hat C$ and by a $(\hat G^{-ab})^2$
term in the action.

\providecommand{\href}[2]{#2}

\end{document}